\documentclass[]{aa}
\usepackage{epsfig}
\usepackage{natbib}
\usepackage{txfonts}

\newcommand{\updown}{\mbox{\boldmath
     $\mbox{}^{\displaystyle \sqcap}\hspace*{-2.6pt}\mbox{}_{\displaystyle
     \sqcup}$}}
\newcommand{\sub}[1]{_{\rm #1}}

\newcommand{\changed}{}

\begin{document}
\title{The stability of spectroscopic instruments: 
A unified Allan variance computation scheme}
\titlerunning{A unified Allan variance computation scheme}
\author{V. Ossenkopf}
\institute{1. Physikalisches Institut der Universit\"at 
zu K\"oln, Z\"ulpicher Stra\ss{}e 77, 50937 K\"oln, Germany
\and
SRON Netherlands Institute for Space Research, P.O. Box 800, 9700 AV 
Groningen, Netherlands
\and
 Kapteyn Astronomical Institute, University of Groningen, PO box 800, 
 9700 AV Groningen, Netherlands
}

\abstract
{The Allan variance is a standard technique to characterise
the stability of spectroscopic instruments used in astronomical
observations. The period for switching between source and 
reference measurement is often derived from the Allan minimum time.
However, various methods are applied to compute the Allan variance 
spectrum and to use its characteristics in the setup of
astronomical observations.
}
{
We propose a new approach for the computation of the Allan variance
of spectrometer data combining the advantages of the
two existing methods into a unified scheme. 
Using the Allan variance spectrum we derive
the optimum strategy for symmetric observing schemes 
minimising the total uncertainty of the data resulting from 
radiometric and drift noise.
}
{The unified Allan variance computation scheme is designed to
trace  total-power and spectroscopic fluctuations within the
same framework.
The method includes an explicit error estimate both for
the individual Allan variance spectra and for the derived
stability time.
A new definition of the instrument stability time 
allows to characterise the instrument even in the case of a
fluctuation spectrum shallower than
$1/f$, as measured for the total power fluctuations in
high-electron-mobility transistors. 
}
{
A first analysis of test measurements for the HIFI instrument
shows that gain fluctuations represent the main cause of instrumental
instabilities leading to large differences between the stability
times relevant for measurements aiming at an accurate determination of the 
continuum level and for purely spectroscopic
measurements. Fast switching loops are needed for a reliable
determination of the continuum level, while most spectroscopic
measurements can be set up in such a way that baseline residuals
due to spectroscopic drifts {\changed are at a lower level} than the radiometric noise.
We find a non-linear impact of the binning of spectrometer channels
on the resulting noise and the Allan time deviating from the description
in existing theoretical {\changed treatments}.
}
{}
\keywords{Methods: data analysis -- Methods: statistical --
Instrumentation: spectrographs}

\maketitle

\section{Introduction}

All radio-astronomical measurements are affected by instabilities
of the gain, the transmission function, and the internal system
noise changing the absolute scale of the measured signal. To
compensate for these drifts, one switches between the
astronomical source and a reference signal -- a known internal
calibrator or a point on the blank sky -- {\changed on} a timescale short
compared to the instabilities. Only the difference signal is used
then \citep[see e.g.][]{Kraus, RohlfsWilson}.

Because of the overheads introduced by switching, the optimum
strategy is not {\changed to switch as fast as possible, but only} as
fast as necessary to suppress the drift noise. Therefore
the characteristic timescales of the instabilities have to be measured.
This can be done in terms of the Allan variance \citep{Allan},
a powerful technique to determine the stability of general radio-astronomical 
equipment, in particular for systems consisting of heterodyne receivers 
and spectrometer backends \citep[e.g][]{Kooi}. 
The Allan variance spectrum can be computed from any sufficiently
long time series of spectrometer dumps taken at fixed instrumental
settings, provided that the integration times for the individual 
dumps are small compared to all instabilities.

When designing the observing modes for HIFI, the heterodyne instrument
of the Herschel Space Observatory, to be launched in 2008 \citep{HIFI}, we
had to develop a reliable Allan variance computation method {\changed for
characterising} the instrument stability so that the precious
observing time is efficiently used in astronomical
observations. The technique developed here can be applied in the same
way to any ground-based telescope when the instability of the
atmosphere is included in the measurement by looking at blank 
sky through the atmosphere.

An optimum algorithm for the computation of the Allan variance spectrum 
should fulfil four requirements: 
\begin{itemize}
\item{}detect and characterise all instabilities and 
their spectral variation across the measured frequency range
\item{}efficiently use the measured data to extract a maximum of 
information from a limited time series 
\item{}provide a measure for the uncertainty of the analysis
itself, i.e. include an error estimate
\item{}be {\changed sufficiently rapid} to allow the use of the Allan variance 
analysis as part of a quick-look analysis of measured data
\end{itemize}

At present, two different algorithms are widely used to compute 
the Allan variance which fulfil the requirements
{\changed given} above only partially. The spectroscopic Allan variance,
as proposed by \citet{Schieder} \citep[see also][]{Kooi}, uses 
only one or two arbitrarily
selected channels to measure total-power or spectroscopic
fluctuations, thus neglecting a large amount of the
measured information. The baseline Allan variance, as
proposed by \citet{Siebertz} \citep[see also][]{WBS}, considers 
all variations across
the spectrometer, but it ignores total-power drifts and
allows no identification of problematic channel ranges within
the spectrum.

Here, we propose a new scheme
which unifies the different approaches into a single mathematical
description and {\changed largely} fulfils the four requirements.
The request for a complete characterisation of the spectral
behaviour of the possible instabilities will be translated 
into the need for a computation of the Allan variance independently for
each backend channel, so that channel by channel variations 
can be detected, the influence of standing wave instabilities
becomes visible, and regions of instabilities across the IF band
can be identified. 

The {\changed requirement} for an efficient data use is
fulfilled by actually using the data from all spectrometer channels
and by taking into account all possible statistically
independent samplings of the temporal behaviour in the analysis.
This efficient {\changed use of data} helps to shorten the actual time needed
{\changed to acquire} the data. Although a complete characterisation
of the instrumental stability requires very long time series of
measurements, the derivation of the Allan variance minimum and the 
drift index of fluctuations at time scales in the order of the
Allan minimum time can be obtained from a 
measurement that lasts only about three Allan minimum times.
Focusing on these two quantities being the actually limiting
factors for the planning and the calibration of astronomical observations
allows to draw significant conclusions also from reasonably short
time series, thus saving observing time. 

Even if the time series of measurements is too short to guarantee
a complete statistical invariance of the data, we can derive
an explicit error estimate for the Allan variance from the
counting statistics of the data taking. With the new definition
of the Allan time proposed here this also provides a direct measure
for the error of the Allan stability time. The request for a
fast implementation is fulfilled by the proposed
convolution schemes for the measured data, either in the 
time domain or in the Fourier domain.

In Sect. 2 we discuss the influence of different data normalisation 
schemes. Sect. 3 summarises our algorithm for the actual Allan
variance computation including the error estimate. Sect. 4
discusses  the best definition of a stability time and the
effect of the binning of spectrometer data and Sect. 5 uses
the results to derive the optimum observing strategy for all 
symmetric astronomical observing schemes. The implications for
asymmetric schemes like on-the-fly mappings will be discussed in
a separate paper. Sect. 6 summarises our results.

\section{Data handling}

\subsection{Normalisation across spectrometers}
\label{sect_normalize}

To perform an instrumental stability analysis a time series of
spectral dumps has to be taken consisting of spectrometer
count rates $c_i(t_k)$, where the index $i$ denotes the
channel number across the backend and $t_k$ gives
the time for the spectral dump with index $k$. The existing
Allan variance analysis tools always assume that the readout dead
time for an individual data dump is negligible relative to the 
integration time between two dumps. Depending on the instrumental
operation, this assumption may be violated in some cases and
we will discuss the effect of dead times below. We have to assume,
however, that every measurement $c_i(t_k)$ {\changed covers} the 
same integration time.

The spectrum of counts per channel $c_i$ on any source
is always dominated by the system bandpass of the instrument which is 
rarely flat {\changed and may be} strongly varying across the 
{\changed covered frequency range}. 
Astronomical data are thus calibrated with respect to this bandpass
using two reference measurements, obtained from the blank sky and 
one or two thermal calibration sources \citep{Kutner}, to deduce 
the actual input signal. To make the stability analysis directly
applicable to astronomical measurements, the same normalisation 
should be used instead of working with raw backend count rates.
However, a full astronomical calibration is neither practical nor
necessary. The bandpass can be approximated by the 
average signal level obtained in the measurement corrected by
the zero level of the instrument.
A useful normalisation of the spectra is thus provided by
\begin{equation}
s_i(t_k)={{c_i(t_k) - z_i } \over {\langle c_i(t_k) -z_i \rangle_k}}
\label{eq_tp}
\end{equation}
where $z_i$ is the zero level of channel $i$ and
the temporal average of each channel is used to normalise
the signal level of that channel. 
This normalisation is also used in the baseline
Allan variance analysis by \citet{Siebertz}. It results in an
approximate equivalence of all backend channels so that 
differences in their mutual behaviour appear on the same
scale. All variations are seen relative to the signal level
so that they can be compared directly to calibration errors
and the radiometric noise level.

\subsection{Spectroscopic versus total power normalisation}
\label{sect_spectroscopic}

Most fluctuations in the amplifiers or other components of
the signal path lead to variations which are constant across
the whole bandpass. Observations which do not switch between 
source and reference on a time scale short compared 
to the corresponding fluctuations will {\changed exhibit}
baseline offsets in the
calibrated data. However, most astronomical heterodyne 
observations are not intended to obtain an accurate
continuum level but for the measurement
of lines on top of a constant baseline which is typically
taken to be zero. In this case, only fluctuations which do
not affect all channels in the same way result in a
degradation of the calibrated astronomical data. Spectroscopic
fluctuations change the mutual response between different 
channels, often seen as ripples or steps in the baseline.

\begin{figure}[ht]
\includegraphics[angle=90,width=8.8cm]{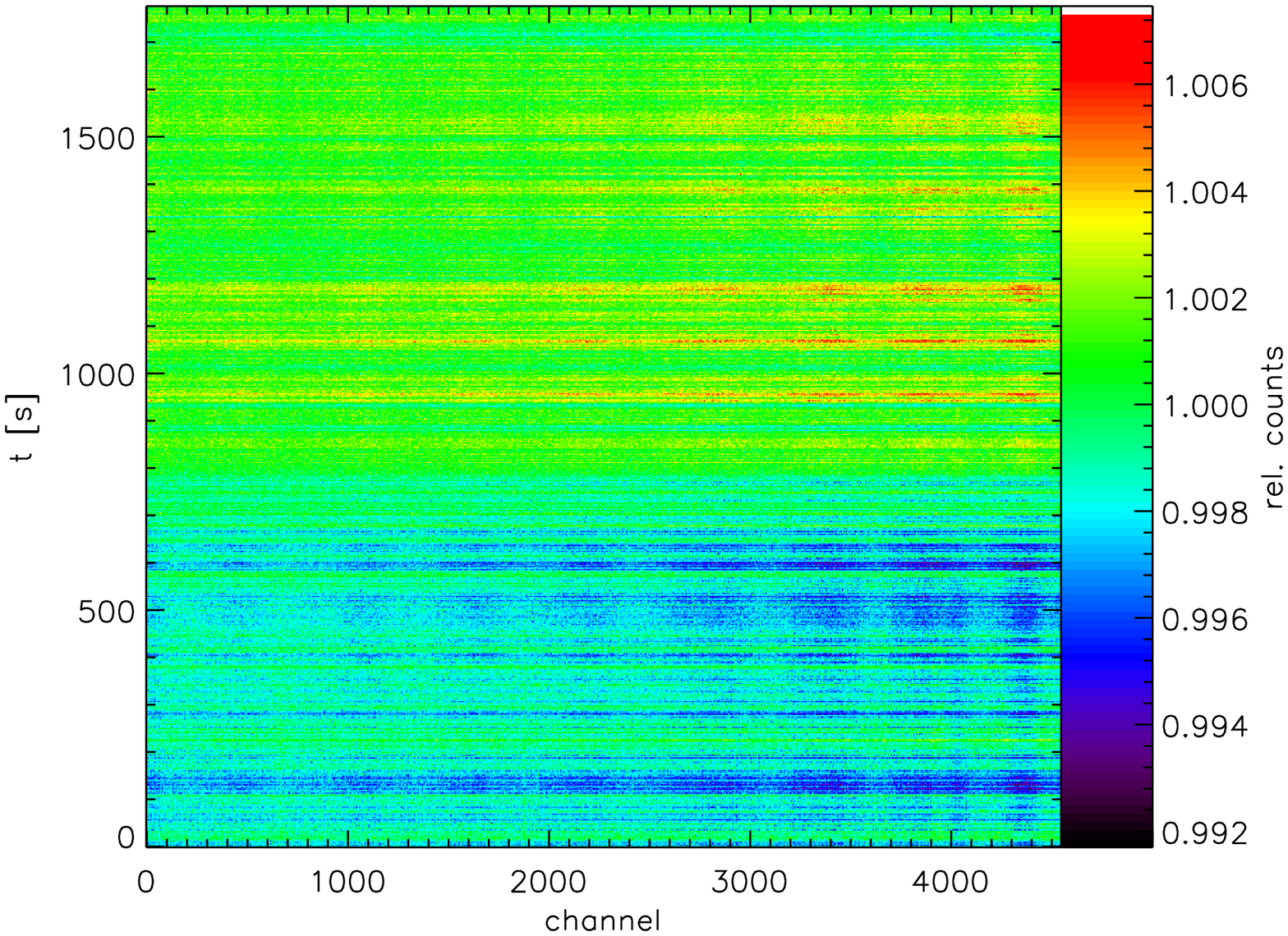}
\vspace{0.1cm}\\
\includegraphics[angle=90,width=\columnwidth]{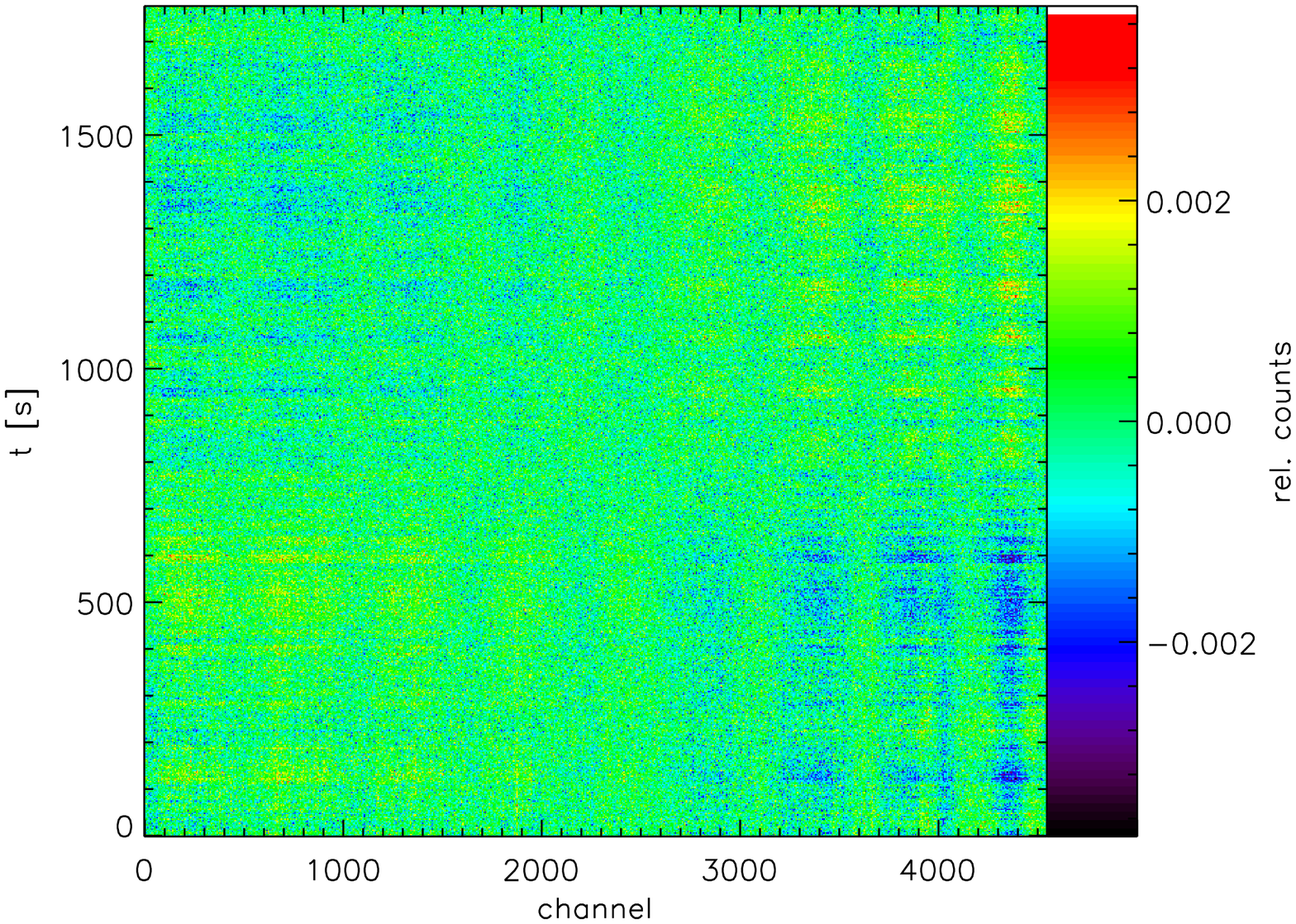}
\caption{Time series of spectrometer data where either the
total-power normalisation (Eq. \ref{eq_tp}, upper plot) or the
spectroscopic normalisation (Eq. \ref{eq_spec}, lower plot) is applied.
The data were taken in stability measurements of the HIFI band 6
IF chain in V-polarisation using the wide band spectrometer 
\citep[WBS,][]{WBS} in 
June 2007. The three WBS subbands covering the available IF bandwidth
{\changed are plotted sequentially, appearing as a single spectrometer with 4556
channels.}
}
\label{fig_timeline}
\end{figure}

Astronomical observations thus call for two different kinds
of instrument stabilities, a total power stability and
a spectroscopic stability. Two different types of Allan variance 
analyses are required. The total power
Allan variance has to trace all instrumental variations. It is 
computed directly from the 
normalised spectra as given in Eq. (\ref{eq_tp}).
The spectroscopic
Allan variance measures only instabilities deviating from a
common gain variation across the whole band. In the original
definition of the spectroscopic Allan variance by \citet{Schieder}
this is accomplished by considering the difference
signal between two channels $i$ and $j$. This approach, however, suffers
from the arbitrariness of the selection of these channels
and the impossibility to distinguish the contributions from either
of the channels.
To avoid these problems, we propose to use the average over the
whole backend to subtract the continuum level fluctuations. We 
extend Eq. (\ref{eq_tp}) by this difference for the spectroscopic 
Allan variance computation:
\begin{equation}
s_i(t_k)={{c_i(t_k) - z_i } \over {\langle c_i(t_k) -z_i \rangle_k}}
- \left \langle {{c_i(t_k) - z_i } \over {\langle c_i(t_k) -z_i \rangle_k}}
\right \rangle_i \;.
\label{eq_spec}
\end{equation}
This spectroscopic normalisation corresponds
to the subtraction of a zero-order baseline in the Allan variance
method proposed by \citet{Siebertz}.
For long time series, the resulting spectroscopic Allan variance 
spectra are also equivalent to the average of two-channel
Allan variance spectra taken over all reference channels $j$,
except for the small contribution from the considered channel $i$
itself to the average which is not used in the two-channel
spectroscopic Allan variance. 

Fig. \ref{fig_timeline} {\changed illustrates} the two normalisations
by plotting a time series of 
spectrometer data $s_i(t_k)$ both for the total power and
the spectroscopic approach. {\changed Most variations are seen to}
occur equally in all channels so that they are mainly visible
in the total power data. However, we also see some differential
variations in terms of gain changes across the full band and
variable standing wave features showing up as weak structures 
in the spectroscopically normalised plot.

HIFI and a number of other receivers cover a large bandwidth 
by combining several sub-spectrometers
with a smaller bandwidth into one large array. In these array
spectrometers, the different subbands will see partially different
signal chains so that they may {\changed show different stability behaviours}.
Each subband has to be characterised individually.
Thus all averages over the channels $i$ were computed only within 
the spectrometer subbands. In this way, we provide the stability
numbers relevant for individual narrow lines, {\changed but we disregard} 
the spectral purity of the full baseline. {\changed The latter} may show steps
between the different subbands due to dynamic platforming, i.e.
a mutual drift of the gain between the subbands. To include these
effects a second spectroscopic normalisation needs to be used where
the average over all channels $i$ in Eq. (\ref{eq_spec}) is computed 
for the full spectrometer. Examples are given in Sect. 4.
In case of sufficient overlap between the subbands, dynamic
platforming can be corrected in the data reduction process
so that we can {\changed restrict ourselves} to the stability analysis for the individual
subbands.

Altogether, each stability measurement characterises two kinds 
of instabilities. When performing the analysis at the normalised
data given by
Eq. (\ref{eq_tp}) we measure the total power stability. This result
has to be used when setting up observations which aim for a determination
of the continuum level, {\changed as is the case for} measurements of absorption lines.
With the spectroscopically normalised data from Eq. (\ref{eq_spec})
we measure the spectroscopic stability, that can be used for
observations which do not aim for an accurate determination of the 
continuum level, e.g. observations of molecular emission lines. 
The spectroscopic stability always exceeds the total power stability
leading to more efficient observing schemes, so that one has to find a
compromise between the need for an accurate continuum level {\changed 
determination} and the request for a high observing efficiency.

\section{Computation of the Allan variance}

\subsection{Convolution schemes}

The computation of the Allan variance consists of a 
convolution of the signal data $s_i(t_k)$ from Eqs. (\ref{eq_tp}) or 
(\ref{eq_spec}) by a Haar wavelet of size $L$ 
\begin{equation}
\updown_{L}=\left\{
\displaystyle
{ 1/L \quad {\rm for}\; -L \le t < 0
\atop \displaystyle
-1/L \quad {\rm for}\; 0 \le t < L }
\atop \displaystyle
0 \quad {\rm everywhere\;else}
\right.
\label{eq_updown}
\end{equation}
and the computation of the variance of the convolved
signal\footnote{Note that the original definition
of the Allan variance is smaller by a factor 1/2. We omit this
factor to allow a direct comparison to the drift error in observations.}$^,$
\footnote{
The variance definition used here is
\begin{displaymath}
\sigma^2 = \left \langle \left (x_k - \left \langle x_k \right \rangle_k
\right)^2 \right \rangle_k = { 1 \over N} \sum_{k=1}^N (x_k - \left \langle x_k \right \rangle_k)^2 \;.
\end{displaymath}
This deviates from the ordinary variance definition
\begin{displaymath}
\sigma\sub{standard}^2 = { 1 \over {N-1}} \sum_{k=1}^N (x_k - \left \langle x_k \right \rangle_k)^2
\end{displaymath}
for small numbers $N$. It has, however, the advantage that it is independent
of the exact way of sampling a given continuous distribution, measuring
only the internal properties of the distribution, as long as the sampling
is dense enough.}:
\begin{equation}
\sigma_{{\rm A},i}^2(L) = \left \langle \left (s_i(t_k)*\updown_L - 
\left \langle s_i(t_k)*\updown_L \right \rangle_k \right)^2 \right \rangle_k \;.
\label{eq_basicallan}
\end{equation}
Plotting the Allan variance $\sigma\sub{A}^2(L)$ as a function of
the filter size $L$ shows the variation of the signal on the scale of the
temporal lag $L$.
Computing the Allan variance by actually convolving the time series
by the $\updown$-filter function for each time step can be considered
a waste of computing time, because neighbouring values
in the convolved time series are no longer statistically independent.
{\changed Independent convolution results are only obtained for 
points separated by a lag larger than the filter size $L$.}
A more efficient method is thus to compute the convolution
integral only for filter settings separated by the filter size
\citep{Schieder}. When we chose filter length $L$ to be
an integer multiple of the step size $\Delta t =t_{k+1}-t_k$, 
i.e. $L=l \times \Delta t$,
we can use the average over a reduced number of points 
to compute the Allan variance
\begin{equation}
\sigma_{{\rm A},i}^2(L) = \left \langle \left (
S_i(K)-S_i(K+1) - \left \langle S_i(K)-S_i(K+1)
\right \rangle_K \right)^2 \right \rangle_K
\label{eq_binning}
\end{equation}
with 
\begin{equation}
S_i(K)= {1 \over l} \sum_{k=Kl+k_1}^{(K+1)l+k_1-1} s_i(t_k)  \;.
\end{equation}
Compared to the full convolution (Eq. \ref{eq_basicallan})
this corresponds to counting only points separated
by $L$ in the convolved function $s_i(t_k)*\updown_L$ when computing
the variance. 
Each data point of the signal contributes twice -- once in the
positive and once in the negative term for the binned signal. 


\begin{figure}[ht]
\centering \includegraphics[angle=90,width=\columnwidth]{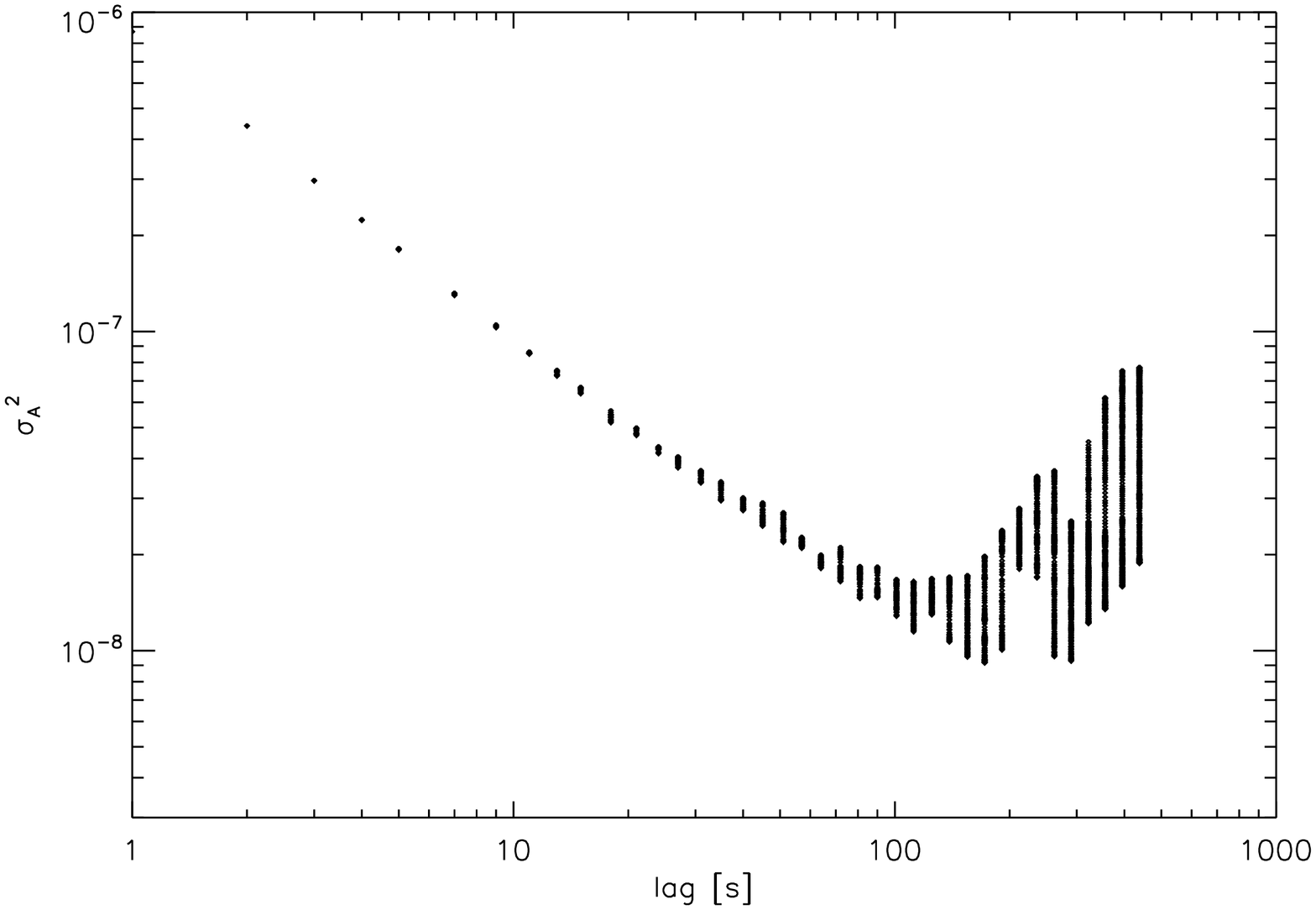}
\centering \includegraphics[angle=90,width=\columnwidth]{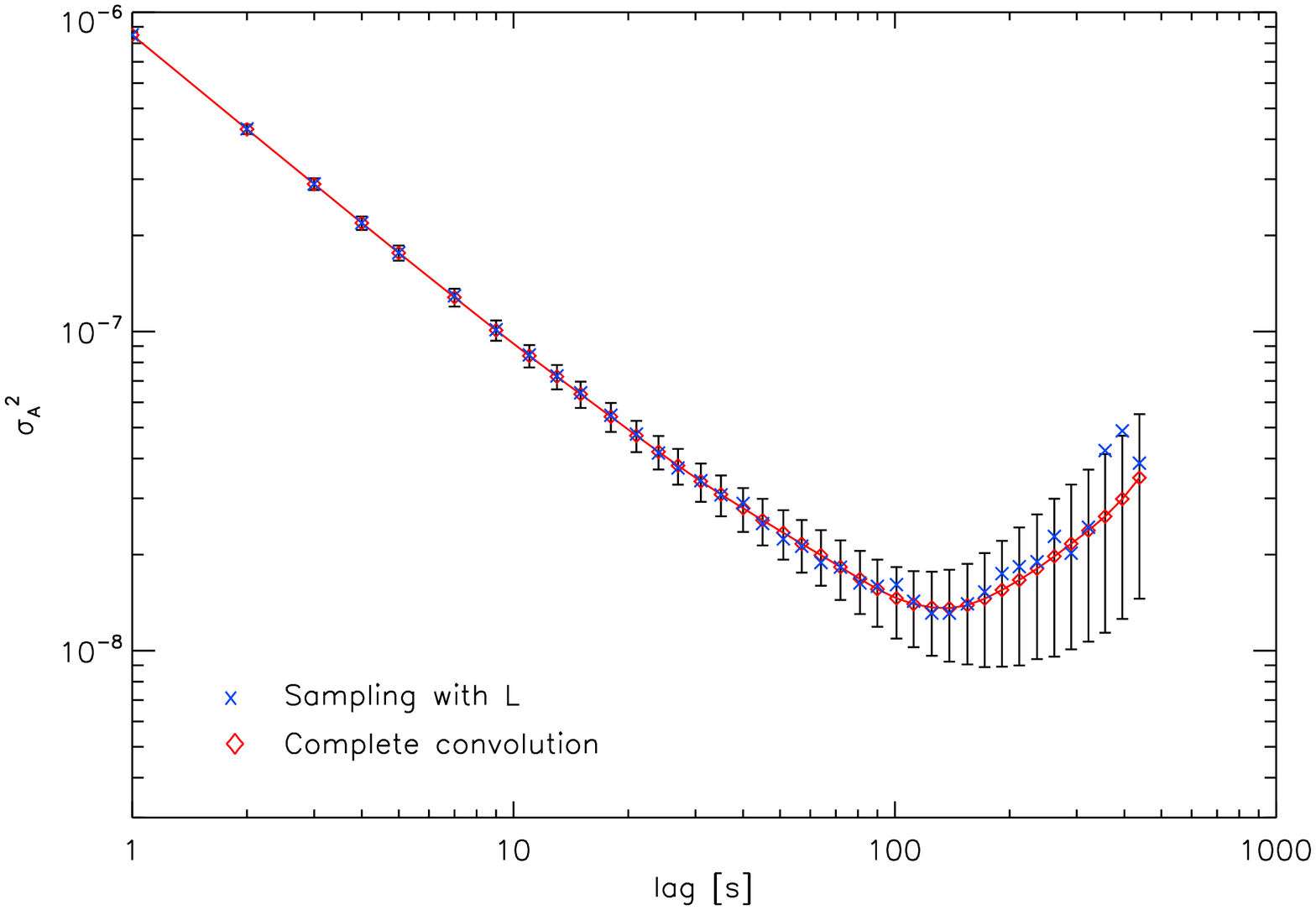}
\caption{Spectroscopic Allan variance spectrum computed
by discrete sampling of the time series in steps of the lag $L$ for
all possible starting points $l_1$ in the time series (upper plot)
compared to the spectrum computed
by full convolution of the time series with the
$\updown$-wavelet (lower plot). The crosses in the lower plot
represent the result from the $L$-sampled computation for $k_1=1$.
The error bars shown for the
results from the full convolution integral are determined by the
statistical uncertainty due to the finite number of data points
(see Sect. \ref{sect_error}).
}
\label{fig_sampling}
\end{figure}

One disadvantage of this method is the loss of information from 
points at the beginning or the end of the time series which are
not covered by full intervals of size $2L$. The discrete sampling
approach necessarily ignores an arbitrary fraction of the data 
at the beginning or end of the time series when the time
series does do not match {\changed integer multiples} of the filter size. 
This is obviously a small effect for small filters, but leads to a
noticeable effect for large filters. 
Moreover, the selection of the start frame $k_1$ for the
convolution is arbitrary. By shifting this through one filter length,
different results are obtained for the Allan variance in case
of a finite statistics in the time series of data. This is demonstrated
in Fig. \ref{fig_sampling}a. For each filter lag $L$ we have computed
the result of Eq. (\ref{eq_binning}) when shifting the start frame
$k_1$ through the full filter length $2L$. We find a large scatter 
in particular for large lags, i.e. a noticeable uncertainty of the 
result {\changed which depends} on the arbitrary selection of this start frame.
We can compare this with the result from the full convolution
of the time series in Fig. \ref{fig_sampling}b. The Allan variance
from the full convolution provides a very smooth curve while the
data from any particular $L$ sampling (shown in Fig. \ref{fig_sampling}b
for $k_1=1$) show a considerable scatter around this curve.
Nevertheless, almost all values fall within the statistical error 
bars computed from the finite size of the data series below.

We can conclude that the brute-force approach of the full convolution
of the time series with the filter function provides the best
results by not neglecting any data and {\changed by} guaranteeing that every feature
in the spectrum is covered by an appropriate filter setting. This
results in smooth curves of the Allan variance spectra facilitating
{\changed fitting and interpreting} by the naked eye. Using today's computer
technology, the convolution is practically possible by means of a 
Fast-Fourier transform for spectrometer time series of up to a few 
thousand steps. 
For longer time series, the numerically simpler approach of the
discrete sampling can be used. The sampling 
in steps of $L$ still provides reliable data
within the statistical uncertainty inherent to the Allan
variance analysis. However, the noticeable error bars for
large lags always have to be taken into account. 
\label{sect_sampling}

\subsection{Error estimate}
\label{sect_error}

To compute the statistical uncertainty of the Allan variance we 
concentrate on this intrinsic uncertainty 
of the method neglecting the error propagation of possible uncertainties
of the measured data into the Allan variance values.

The statistical error results from the sampling error when scanning
a continuous distribution $f(t)$ by taking $N$ data values at discrete
randomly selected points $K$. It is well known that the uncertainty
in the determination of the average value of the sampled distribution is
given by a Poisson counting error and the variance of the
distribution
\begin{equation}
\delta \langle f \rangle_K = 
\sqrt{\langle (f-\langle f\rangle_t)^2 \rangle_t \over N}  \;.
\label{eq_poisson}
\end{equation}

Equivalently, one can derive the uncertainty of the measured variance of the
distribution caused by the discrete sampling as
\begin{eqnarray}
\delta \langle{ (f-\langle f\rangle_K)^2} \rangle_K &=&
\sqrt{
\langle (f-\langle f\rangle_t)^4 \rangle_t - 
\langle (f-\langle f\rangle_t)^2 \rangle_t^2 \over
N} \nonumber \\
& = &
\langle (f-\langle f\rangle_t)^2 \rangle_t \sqrt{ Kur-1
\over N} 
\end{eqnarray}
where $Kur$ denotes the kurtosis of the distribution, characterising
its fourth moment. Gaussian distributions exhibit a kurtosis value of 3.
Exponential distributions show $Kur=6$. Thus the relative accuracy of the
measured variance of a distribution depends mainly on the number
of points used to sample the distribution. The kurtosis measuring the
relative strength of the wings of the distribution weighs this
counting error by a factor of a few.

It is obvious that these principles apply as well to the determination
of the Allan variance. As discussed above, each distribution of filter-convolved
data values can only be sampled in steps of $L$ to obtain
statistically independent values. If the data represent a random
series, {\changed sampling in uniform} time steps corresponds to a random sampling
of the distribution of data values so that the equation above can be applied.
This assumption is not always fulfilled in the measurement
of drift processes but it is in general justified when sufficiently
long time series are measured. Thus we can estimate the statistical
error of the
Allan variance for a time series with the length $N \times L$ by
\begin{equation}
\delta \sigma\sub{A,i}^2(L) =\sqrt{
\langle Z_{i,k}^4(L) \rangle_k - 
\langle Z_{i,k}^2(L) \rangle_k^2 \over
N} 
\label{eq_deltaerr}
\end{equation}
with
\begin{equation}
Z_{i,k}(L) = s_i(t_k)*\updown_L -\langle s_i(t_k)*\updown_L\rangle_k \;.
\end{equation}
These error bars {\changed are} plotted in Fig. \ref{fig_sampling}. 
When comparing the plots in Fig. \ref{fig_sampling}, it is remarkable
that the scatter obtained while varying $k_1$ gives a very good
match to the statistical uncertainty shown as error bars of the 
convolved data.

\subsection{How to characterise a full spectrometer?}

\begin{figure}
\centering \includegraphics[angle=90,width=\columnwidth]{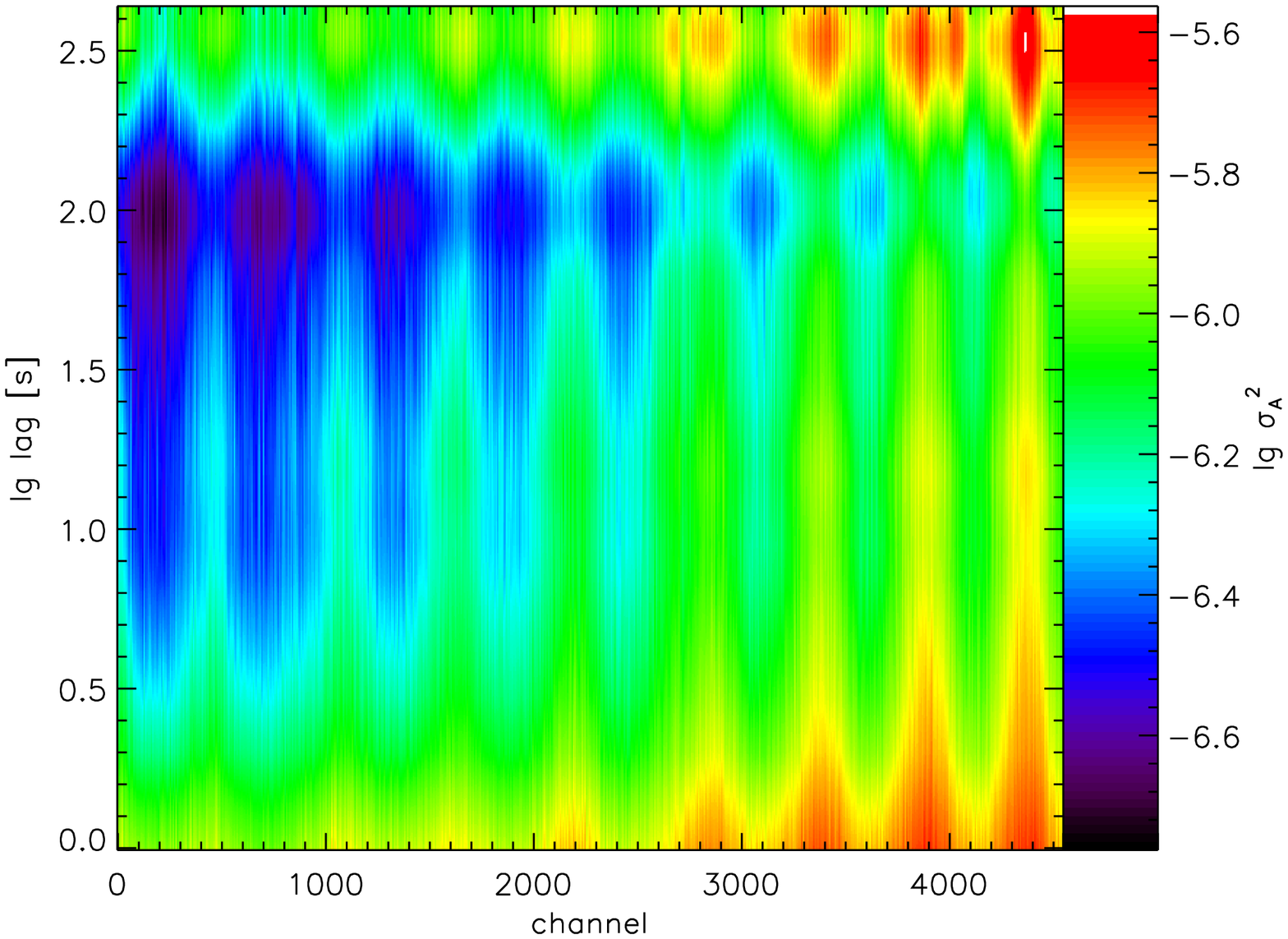}
\vspace{0.1cm}\\
\hspace*{-0.8cm}\includegraphics[angle=90,width=8cm]{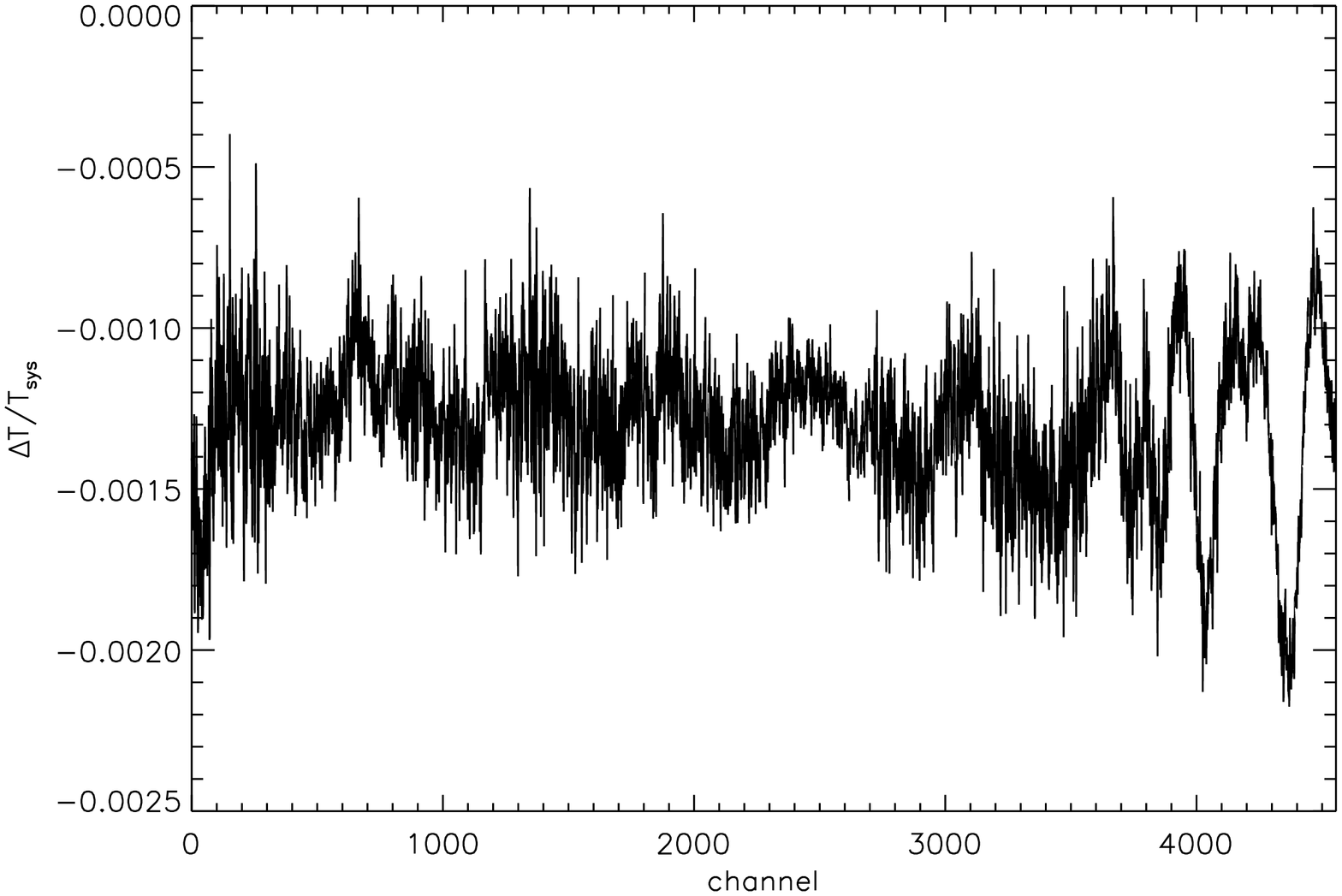}
\caption{Channel-by channel total-power Allan variance spectra
for the test measurement described in Fig. \ref{fig_timeline} (upper plot).
The colour coding shows the logarithm of the Allan variance.
The lower plot shows the difference of two spectra obtained by
averaging over two {\changed consecutive} 500s intervals.
}
\label{fig_allansurface}
\end{figure}

All computations so far were restricted to a time series of
arbitrary data, i.e. an individual spectrometer channel $i$, but
they ignored that a full instrument consists of thousands of 
spectrometer channels which
are partially, but not completely independent. 

The first approach to characterise the full spectrometer
is the computation of the Allan variance channel by channel and the
visualisation of the result in a three-dimensional plot. 
The result of such a channel-by channel analysis is demonstrated in 
the upper part of Fig. \ref{fig_allansurface} showing the total-power
Allan variance in logarithmic units for the HIFI test measurement 
from Fig. \ref{fig_timeline}. Three subbands of the wide band 
spectrometer cover the full spectrum; their boundaries are
visible at channels 860 and 2708. We clearly see that the instability
of the system is dominated by a kind of standing-wave pattern
superimposed on a large-scale trend of higher stabilities
at {\changed lower frequencies (smaller channel numbers)}. We find a pattern of
alternating regions of {\changed more and less stable} channels. This can
be interpreted in terms of the baseline of example spectra.
The lower plot in Fig. \ref{fig_timeline} shows the difference
of two spectra obtained by averaging over two adjacent 500~s 
intervals. We recognise a kind of standing-wave baseline distortion
with a period matching the features visible in the Allan variance
spectra. {\changed The amplitude of this baseline distortion
grows towards higher channel numbers where we find larger
values of the Allan variance at the upper boundary 
of the three-dimensional plot corresponding to $L \approx 500$~s.}

From this plot it is obvious that variations of the instrumental
behaviour across the spectrometer have to be considered.  This
provides the full information, but has the
practical {\changed disadvantage} that a surface plot is 
more difficult to {\changed interpret} by eye than a two-dimensional
plot, that it is not possible to include information about the
error bars in the plot, and that for the optimisation of the
observing strategy we have to reduce the result to
a few numbers. To get a rough feeling for the spectral behaviour
one can also look at the difference baseline plot, however,
the corresponding interval has to be selected arbitrarily,
missing most of the statistics of the measurement, so that
a single baseline can never provide all the information contained
in the channel-by-channel Allan variance plot.

For the derivation of constraints for the observing strategy it
seems plausible to characterise the whole instrument by the 
properties of the worst, i.e. most {\changed unstable} channel. This is
the most reliable approach also applicable to observations where
the measurement in all channels is equally important, e.g. in
frequency surveys of rich emission spectra with hundreds
of lines per spectrum.
However, it does not take into account that in most observations
the observed lines cover only a very small fraction of the whole
spectrometer output whereas bad channels are
typically concentrated towards the edges of the IF band. Then, an 
average Allan variance spectrum is more appropriate. 

One can consider three different ways of averaging. When starting
from the channel-by-channel Allan variance analysis an average
Allan variance spectrum is given by
\begin{equation}
\sigma\sub{A}^2(L) = \left \langle \left \langle \left (
s_i(t_k)*\updown_L - \left \langle s_i(t_k)*\updown_L
\right \rangle_{k} \right)^2 \right \rangle_{k}
\right \rangle_i \;.
\end{equation}
This approach corresponds to averaging the Allan variance spectrum
obtained from two channels following the method by \citet{Schieder}
over all pairs of channels.

A second possible approach is used by the baseline Allan 
variance method by \citet{Siebertz}. The variance within the
convolved spectrum for each time step is considered and in a second step
the average over all time steps is performed. We can write this as
\begin{equation}
\sigma\sub{A}^2(L) = \left \langle \left \langle \left (
s_i(t_k)*\updown_L - \left \langle s_i(t_k)*\updown_L
\right \rangle_i \right)^2 \right \rangle_i
\right \rangle_{k} \;.
\end{equation}
This kind of averaging is not able to monitor total-power variations
because they enter the data $s_i(t_k)*\updown_L$ in the same way as 
$\left \langle s_i(t_k)*\updown_L \right \rangle_i$ so
that they are subtracted and removed from the Allan variance spectrum.
Therefore, {\changed the variance} can only measure the spectroscopic Allan variance
of an instrument.

The third option is to determine the variance relative to the grand
average of the normalised and convolved spectra from the whole
data field
\begin{equation}
\sigma\sub{A}^2(L) = \left \langle \left (
s_i(t_k)*\updown_L - \left \langle s_i(t_k)*\updown_L
\right \rangle_{k,i} \right)^2 \right \rangle_{k,i} \;.
\end{equation}

\begin{figure}
\centering \includegraphics[angle=90,width=\columnwidth]{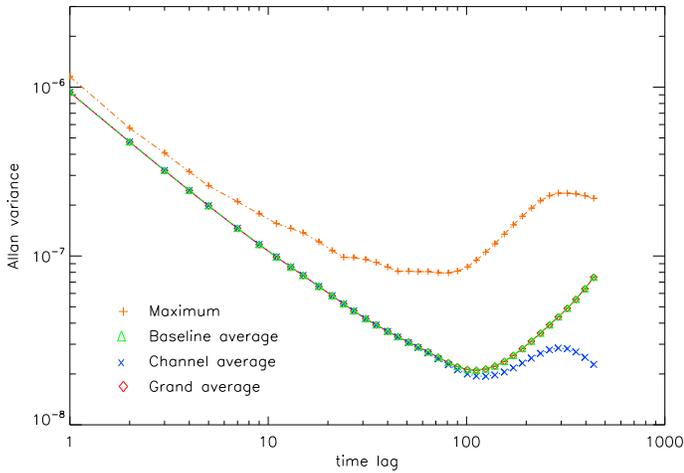}
\caption{Comparison of the four possible ways to characterise
a  full spectrometer band by a single Allan variance spectrum.
Spectroscopically normalised data from the central subband
shown in Fig. \ref{fig_timeline} have been used.
The results when averaging the baseline Allan variance plots
and when determining the variance relative to the grand average
including simultaneously the channel scale and the time scale are
so close that the curves are hardly distinguishable by eye.
\label{fig_averaging}}
\end{figure}

The result for the spectroscopically normalised time series of
the HIFI test measurement shown in Fig. \ref{fig_timeline} are
demonstrated in Fig. \ref{fig_averaging}.
At all time lags below 100~s the different averages show 
identical Allan variance spectra with a slope of $-1$ 
characteristic for white radiometric noise.
In contrast, the selection of the worst channel shows that 
this experiences some additional fluctuations even at the scale of 1~s
and a drift noise growing to the amplitude of the radiometric
noise already after 20~s. Consequently, {\changed the relevant
stability times are} very short if one has to guarantee that the drift 
contribution remains small for each individual channel. The 
vast majority of all spectrometer channels behaves {\changed in a 
much more stable manner}.

Comparing the average of the baseline Allan variance with the
Allan variance using the grand average in the difference shows
very similar spectra which are hardly to
distinguish by eye, i.e. the variation of the convolved data
relative to the average
over the spectrum at a given time step is almost identical to
the variation relative to the global average. In contrast, the 
average of the channel Allan variance spectra is always smaller
at large lags, i.e. the variation of the convolved data relative
to the average in the corresponding channel is always smaller
than the variation relative to the global average. 
This can be understood from the 
the bandpass normalisation and the spectroscopic normalisation of
the signal $s_i(t_k)$ discussed in Sect. \ref{sect_normalize}.
The bandpass normalisation guarantees that the average 
$\left \langle s_i(t_k) \right \rangle_{k}$ of any channel $i$
is identical to the global average $\left \langle s_i(t_k)
\right \rangle_{k,i}$ whereas the spectroscopic normalisation 
leads guarantees the identity 
$\left \langle s_i(t_k) \right \rangle_{i} = \left 
\langle s_i(t_k) \right \rangle_{k,i}$ for each time step.
For spectroscopically normalised data, the convolution with 
the Allan filter hardly changes the average of the individual spectra,
$\langle s_i(t_k)*\updown_L  \rangle_{i}
\approx  \langle s_i(t_k)*\updown_L  \rangle_{k,i}$
for each time step $i$, but for the individual channels, the 
convolution leads to offsets of $\langle s_i(t_k)*\updown_L \rangle_{k}$ 
from the global average when they have different
trends {\changed which do not exactly cancel}. For 
spectroscopically normalised data and sufficiently long time series 
{\changed in which all drifts do statistically cancel},
all three averaging methods should provide the same results.

\begin{figure}[ht]
\centering \includegraphics[angle=90,width=\columnwidth]{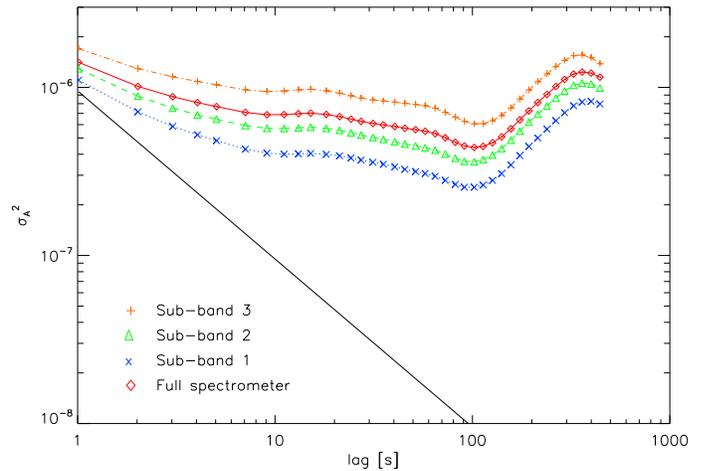}
\caption{Average total power Allan variance spectra for the
different WBS subbands computed for the time series of 
data presented in Fig. \ref{fig_timeline}.
}
\label{fig_tpaverage}
\end{figure}

For practical application, the method should be chosen
which is most adequate {\changed for} the need of astronomical observations
to control the drift error in measured data. Basically all spectroscopic 
observations are interested in the shape of a whole spectrum 
which is more than a bunch of independent channel data. 
Drift errors, showing up as irregular baseline 
distortions, can result in a clear degradation of the scientific value
of the data even if the magnitude of the distortions does not exceed
the radiometric noise in the data. To take the mutual relation of 
drift contributions across the spectrum into account we have
to use the average of the baseline Allan variance or
the Allan variance relative to the grand average. Because the
baseline Allan variance is not able to characterise total power drifts
we propose to always use the "grand average" method. It
combines the advantage of the baseline Allan variance method
reflecting the observer's view on
spectroscopic data with the possibility {\changed of analyzing} the drift
behaviour of an instrument including total-power variations.
However, we want to stress again that the characterisation of the
instrument by a single Allan variance spectrum is only justified
if an inspection of a plot like Fig. \ref{fig_allansurface} has
{\changed revealed} that no strong deviations of the drift behaviour across the
spectrometer occur. 

An intermediate level of analysis can be provided for 
array-spectrometers, like the HIFI-WBS, by characterising
each spectrometer subband individually (see Sect. \ref{sect_spectroscopic}).
Fig. \ref{fig_tpaverage} shows the resulting total-power
Allan variance spectra for the HIFI laboratory measurement
already used for demonstration in Fig. \ref{fig_allansurface}.
The solid line represents the limit of pure radiometric noise.
{\changed The trend of channels of higher number (frequency) 
having lower stability (Fig. \ref{fig_allansurface}) is clearly seen.}
For the full spectrometer
we obtain a kind of intermediate behaviour. Comparing the plot
with Fig. \ref{fig_averaging} shows that the total-power Allan variance
values suffer from much stronger drift contributions than 
the spectroscopically normalised values, with a noticeable
drift already {changed visible} at very small lags. At all time lags below 100s,
the spectrum shows a typical slope of about $-0.3$. This is in agreement
with stability measurements of the high electron mobility
transistors used in the receiver amplifiers \citep{Whyborn}, 
indicating that the total-power stability of the overall system 
is mainly determined by their gain fluctuations.

\section{Interpretation of Allan variance spectra}

\subsection{Comparison to radiometric noise}

\begin{figure}
\centering \includegraphics[angle=90,width=\columnwidth]{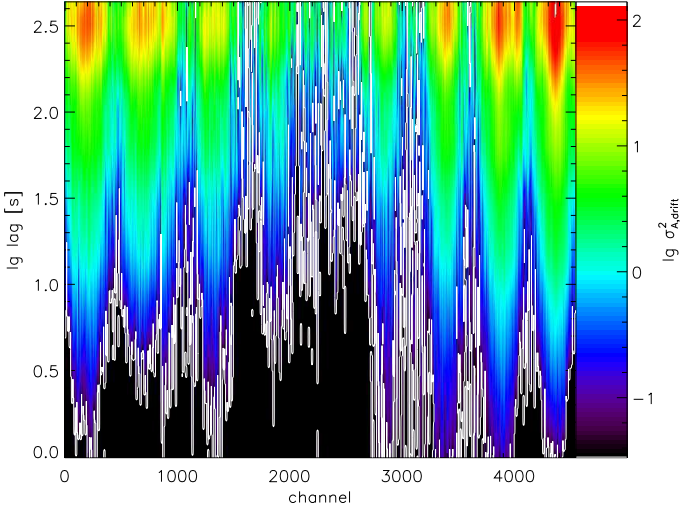}
\caption{{\changed Radiometrically normalized channel-by-channel 
spectroscopic drift Allan variance spectra, 
$\sigma\sub{A,drift}^2(L)=B\sub{Fl} L \sigma_{A}^2(L)/2 -1$,
for the average fluctuation bandwidth of $B\sub{Fl}=2.1$~MHz. 
The color bar gives the logarithm of this variance. White
contours indicate parts of the spectrum where the measured
variations drop below the average radiometric noise level, i.e.
where the normalized spectra fall below zero.} The data 
from Fig. \ref{fig_timeline} are used
so that the figure can be compared to the 
Allan variance plot in Fig. \ref{fig_allansurface}.
\label{fig_sallansurface}}
\end{figure}

The Allan variance always contains a combination of radiometric
and drift fluctuations. They can be separated based on
their different spectral
characteristics. Fluctuations with a $f^{-\alpha}$ power
spectrum show up in the Allan variance as $L^{\alpha-1}$ spectra
\citep{SchiederKramer}. The superposition of white radiometric noise
($\alpha=0$) and a power law drift noise with an arbitrary spectral
exponent $\alpha$ gives an Allan variance spectrum
\begin{equation}
\sigma\sub{A}^2(L)= {2 \over B\sub{Fl} L} + A L^{\alpha-1} \;.
\label{eq_spectralallan}
\end{equation}
Here, $B\sub{Fl}$ is the fluctuation bandwidth
per backend channel determined by the power spectrum of the
noise \citep{Kraus} and $A$ characterises the amplitude of
the instrumental drift. 

Subtracting the radiometric noise
contribution, $2/(B\sub{Fl} L)$, from the Allan variance spectrum
isolates the drift contributions. Moreover, it is useful
to normalise the drift contribution by the radiometric noise
because all observations will aim for data where
the error due to instrumental drifts is small compared to the
radiometric error of the observation and the radiometric noise
can be easily calculated. We obtain the
normalised drift Allan variance as
\begin{equation}
\sigma\sub{A,drift}^2(L)={B\sub{Fl} L \over 2}  \sigma_{A}^2(L)-1
={B\sub{Fl} A \over 2} L^\alpha \;.
\end{equation}
{\changed This} describes the actual impact of the drift noise for an 
astronomical observation, {\changed as is demonstrated} in 
Fig. \ref{fig_sallansurface} for the spectroscopic channel-by-channel
Allan variance of the data from Fig. \ref{fig_timeline}.
We find some regions where the
drift contribution seems to be negative, indicated by the
white contours. This is due to the fact that the fluctuation
bandwidth is not completely constant across the spectrometer
and when subtracting the radiometric noise for an average
fluctuation bandwidth, we create slightly negative numbers
for channels that have a fluctuation bandwidth that is eventually
slightly larger than the average. The effect is always less
than a few percent, so that we neglect it here. We recognise 
again the pattern of standing-wave like drift
contributions, but the large scale gradient that
was visible in the total-power Allan variance from Fig.
\ref{fig_allansurface} {\changed is absent}. The drift noise is strongly variable
across the spectrometer. At $L=300$s it hardly exceeds the
radiometric noise in the most stable regions while it exceeds
it by a factor $>100$ in the most unstable regions of the
spectrum.

We can use the knowledge {\changed of} the shape of the radiometric
noise contribution to derive the instrumental stability
also from imperfect measurements not covering a continuous
time series. If the spectrometer readout
adds a significant dead time between individual integrations,
Eq. (\ref{eq_spectralallan}) does not hold any more because
the radiometric noise is determined by the fraction of $L$
used for the integration, $L\sub{int}=L-L\sub{dead}$, while
the drift is still determined by the total time lag between
two measurements $L$. We can correct for these dead times
as long as the Allan variance at the first lag 
$\sigma\sub{A}^2(L_1,L\sub{int,1})$
is still purely radiometric or if we know the level of the radiometric
noise from independent computations. If we scale the radiometric
contribution by the factor $L\sub{int}/L$, we obtain again
a self-consistent Allan variance spectrum, depending only on
the filter size $L$:
\begin{equation}
\sigma\sub{A}^2(L)=\sigma\sub{A}^2(L,L\sub{int}) - 
\sigma\sub{A}^2(L_1,L\sub{int,1}) \times L\sub{dead,1}/L \;.
\label{eq_deadtimes}
\end{equation}
This corrected spectrum can be used in the same way as the
spectrum from perfect measurements to derive the Allan time.

\label{sect_drifttoradio}

\subsection{Definition of the stability time}

The Allan time is used to quantify the lag
at which the Allan variance spectrums changes from being dominated
by radiometric noise to being dominated by the instrumental drift.
The traditional definition 
uses the minimum of the Allan variance spectrum. At smaller
lags the fluctuations are dominated by the radiometric noise that
drops with the integration time given by the Allan filter size,
at larger lags the
drift terms dominate resulting in an increase of the fluctuations
with filter size. This approach, however, is only applicable if
the drift follows a usual spectral characteristics with an
index $\alpha> 1$. Otherwise no minimum is formed. Figure
\ref{fig_tpaverage} showed an example with an $L^{-0.3}$ dependence
of the Allan variance corresponding to a $1/f^{0.7}$ 
characteristics of the fluctuations and we have seen numerous
other examples with a drift behaviour close to $1/f$ noise
leading to a very flat Allan variance spectrum. 

The use of the Allan minimum time $t\sub{A}$ 
thus has two disadvantages: \\
{\bf i)} for drift noise
shallower than $1/f$ the Allan variance has no minimum although
it is still a good measure for the stability of the system.\\
{\bf ii)} a small uncertainty in the spectral index of the drift
function can lead to large shifts of the minimum making it very
difficult to derive an accurate error estimate for the Allan 
minimum time from the statistical uncertainty
of the Allan variance.

Therefore, we propose another definition of the Allan time $t\sub{A}'$
based on the normalisation discussed above:
{\it $t\sub{A}'$ is the lag {\changed for which} the drift contribution to
the total uncertainty equals the radiometric contribution, i.e.
$\sigma\sub{A,drift}^2(t\sub{A}')=1$}. This means that
at $t\sub{A}'$ the total Allan variance amounts to twice the
radiometric Allan variance.  For a drift noise following
a $1/f^{2}$ spectral dependence, the new Allan time $t\sub{A}'$
agrees with the Allan minimum time $t\sub{A}$.

The new Allan time definition has two practical disadvantages
compared to the Allan minimum time:\\
{\bf i)} It is easier to determine the minimum of the Allan 
variance by the naked eye than determining the point where it
deviates by a factor 2 from the radiometric line.\\
{\bf ii)} Because most existing measurements characterise the
instrument stability by the Allan minimum time, a comparison with
their results requires an additional translation step.
The relation between both Allan times can be computed
from Eq. (\ref{eq_spectralallan}) as
\begin{equation}
t\sub{A}'= (\alpha-1)^{1/\alpha} t\sub{A} \;.
\end{equation}

However, we can easily compute the uncertainty of the new 
Allan time $t\sub{A}'$ using the uncertainty of the Allan variance 
(Eq. \ref{eq_deltaerr}) by
\begin{equation}
{ \delta t\sub{A}' \over t\sub{A}' } = { 1 \over |\alpha|}
{ \delta \sigma\sub{A}^2(t\sub{A}') \over \sigma\sub{A}^2(t\sub{A}') } \;.
\end{equation}
This error remains finite even in case of $1/f$ noise whereas
it diverges for the traditional Allan minimum time.

\subsection{Binning of spectra}

In astronomical data analysis it is common to use the average
of several channels of a spectrometer to reduce the observational
noise if the fixed resolution of the spectrometer is higher
than the resolution needed to deduce physical parameters
from the observed spectra. An example are
observations of the [C{\sc II}] line at 1.9~THz with the HIFI-WBS
{\changed in which} the astronomer asks for a frequency resolution of about 1~km/s,
corresponding to 6.3~MHz while the native spectrometer resolution 
is 1.1~MHz. It is a common misconception to assume that the
noise in the spectra will drop by a factor $\sqrt{1/11}$ if the 11 channels
are co-added that provide the effective resolution of 6.3~MHz.

For a correct treatment, the correlation between 
neighbouring channels has to be taken into account. The spectral
correlation of the noise leads to a fluctuation bandwidth that
differs from the channel spacing. If the autocorrelation
function (ACF) of the spectrometer is known, the effective fluctuation bandwidth
can be computed as a function of the binning width $n\sub{bin}$, following the 
formalism provided in Appendix A of \citet{SchiederKramer},
\begin{equation}
B\sub{Fl}(n\sub{bin})=n\sub{bin} \Delta \nu\sub{channel} \times
{1 + 2 \sum_{m=1}^{\infty}g_m \over 1 + 2 \sum_{m=1}^{n\sub{bin}-1}
\left[1-m/(n\sub{bin}-1)\right]g_m}
\label{eq_acf}
\end{equation}
where $\Delta \nu\sub{channel}$ denotes the channel spacing and $g_m$ represent
the values of the ACF at discrete channel shifts $m$. 
One can see that the native fluctuation bandwidth of the spectrometer 
($n\sub{bin}=1$) always exceeds the channel spacing by 
the symmetrically integrated autocorrelation function, for the HIFI WBS
this is a factor of almost three. The ACF
has typically only a few non-vanishing coefficients $g_m$, so that
{\changed for large values of $n\sub{bin}$ (large total binning widths)}
$B\sub{Fl}$ approaches $n\sub{bin} \Delta 
\nu\sub{channel}$. Consequently, binning does not reduce the radiometric
noise like $\sqrt{1/n\sub{bin}}$ relative to the noise at the native
resolution, but in the limiting case of large bin widths the noise is 
still about a factor $\approx \sqrt{3}$ higher.

\begin{figure}[ht]
\centering \includegraphics[angle=90,width=\columnwidth]{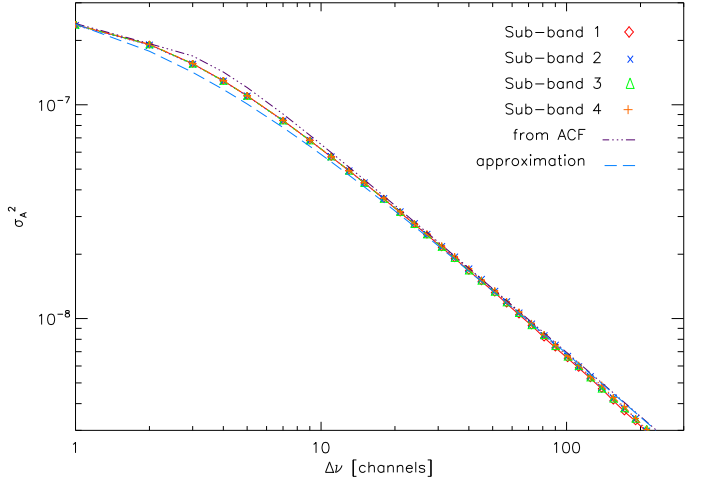}
\caption{Spectroscopic Allan variance at $L=5$~s for stability measurements of
the HIFI WBS (V-polarisation) as a function of the channel binning width.
The difference between the four WBS subbands is too small to be 
visible by eye. The dash-dotted line shows the theoretical curve
computed from the measured autocorrelation function following 
\citet{SchiederKramer}. The dashed line represents the simple
approximation with an additive constant.}
\label{fig_bfluct}
\end{figure}

In Fig. \ref{fig_bfluct} we show this behaviour by computing the Allan 
variance for short lags in a stability measurement of the HIFI 
wide band spectrometer when applying a varying channel binning to the
data. The dash-dotted line represents the curve given by Eq. (\ref{eq_acf})
using laboratory data for the ACF measured by
tracing a tunable narrow line source. We find a good agreement
between the theoretical curve and the measured Allan variance data
showing that we deal with pure radiometric noise at short lags, but 
a deviation by up to 10\,\% at $n\sub{bin} \approx 3\dots 7$. The
reason for this deviation is not clear. An imperfect line source
used when measuring the ACF might lead to an overestimate of $g_m$
for higher $m$ at cost of $g_1$, resulting in this kind
of deviation. Taking the typical error bars of the
Allan variance (Sect. \ref{sect_error}) and the natural variation 
of the fluctuation bandwidth across a spectrometer (Sect.
\ref{sect_drifttoradio}) into account, this small deviation can be
completely neglected in the following.

To quantitatively describe the observed behaviour of the fluctuation 
bandwidth it is possible to use an even simpler approximation
consisting just of a linear dependence and an additive constant
for the native fluctuation bandwidth:
\begin{equation}
B\sub{Fl}(n\sub{bin})=B\sub{Fl}(1)+ (n\sub{bin}-1) \Delta \nu\sub{channel} \;.
\label{eq_simple}
\end{equation}
This is shown as the dashed line in Fig. \ref{fig_bfluct}. We find
again a small deviation at intermediate bin sizes, but an overall
reasonable agreement, so that we can use this approximation for
fast computations.

With known fluctuation bandwidth for a particular channel
binning we can compute the corresponding Allan time by determining the 
ratio between drift noise and reduced radiometric noise. 
It is usually assumed \citep{SchiederKramer, Kooi2006} 
that the binning only reduces the radiometric 
noise while the instrumental drift itself is not affected.
Resolving Eq. (\ref{eq_spectralallan}) for $t\sub{A}'$ gives
\begin{equation}
t\sub{A}'=\left( 2 \over A B\sub{Fl} \right)^{1/\alpha}
\label{eq_atotarelation}
\end{equation}
showing that the Allan time shifts to smaller lags
when increasing the fluctuation bandwidth by
\begin{equation}
t\sub{A}'(n\sub{bin})= \left ({ B\sub{Fl}(1) \over B\sub{Fl}(n\sub{bin})}
\right )^{1/\alpha} t\sub{A}' (1) \;.
\label{eq_binningta}
\end{equation}
This relation is identical to the corresponding relation for the
traditional Allan minimum time \citep{SchiederKramer}.

\begin{figure}[ht]
\centering \includegraphics[angle=90,width=\columnwidth]{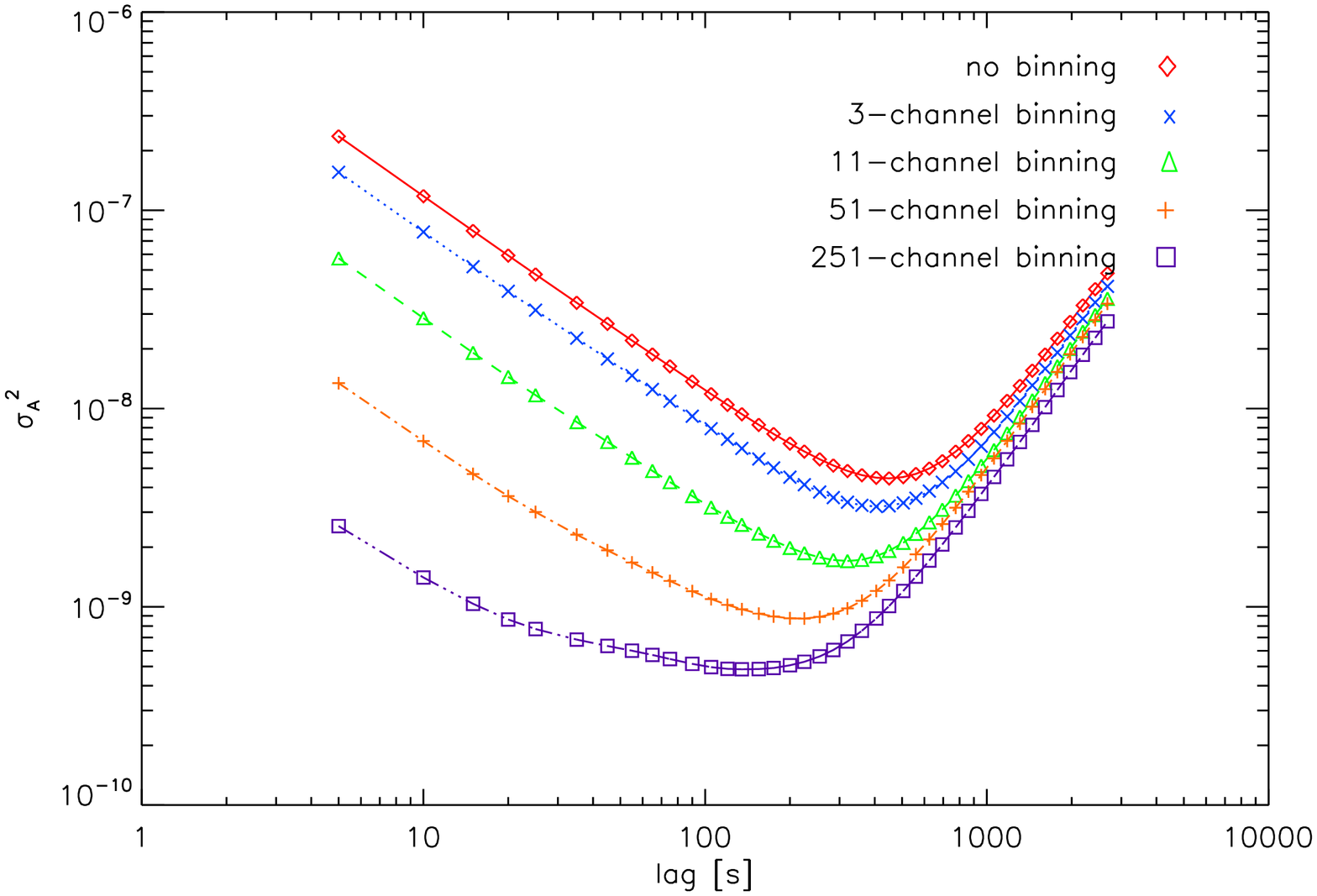}\\
\centering \includegraphics[angle=90,width=8.8cm]{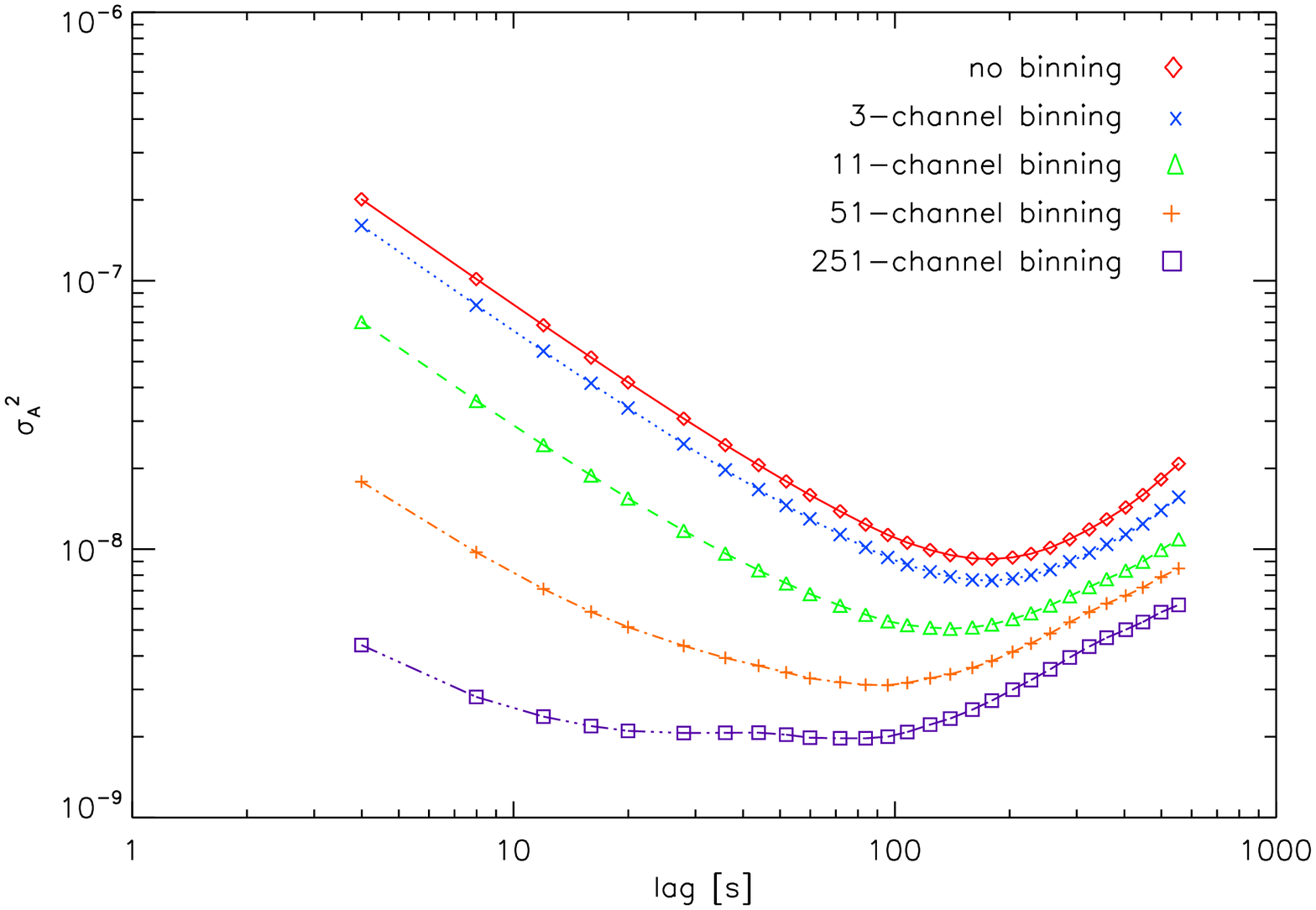}
\caption{Spectroscopic Allan variance spectra characterising the stability of
subband 2 of the HIFI wide band spectrometer using a stable noise source 
(upper plot) and using the whole system (HIFI band 3a, lower plot) for 
{\changed five different levels of channel binning}. }
\label{fig_binningdrift}
\end{figure}

Figure \ref{fig_binningdrift} shows the impact of binning on the
Allan variance spectra for data
taken in a stability test of the spectrometer (upper plot)
and for a test of the full system including mixer, local
oscillator, and IF amplifier chain (lower plot). We show the
spectroscopic Allan variance spectra for four different binnings
and the native resolution. The data characterising the spectrometer
{\changed alone} perfectly follow the theoretical assumption of a constant
drift contribution. Changing the binning width only changes the
radiometric noise so that Eq. (\ref{eq_binningta}) applies.

For the whole system, however, we find a significant change of the 
drift contribution with the binning width, indicating that the
assumption of a constant drift term is in general not applicable. 
We notice two effects:\\
{\it i)} the system is affected by some spectrally correlated fluctuations,
like the {\changed unstable} standing wave patterns discussed before, which are
reduced by spectral binning.
This leads to a reduction of the drift term and consequently to
an increase of $t\sub{A}'(n\sub{bin})$ relative to the value
predicted by Eq. (\ref{eq_binningta}).\\
{\it ii)} the overall system is always affected by some $1/f$ noise.
When steeper drift noise and the radiometric contribution {\changed are}
sufficiently
reduced by binning, the $1/f$ noise starts to dominate the Allan 
variance. This reduces the Allan time relative to the value
predicted by Eq. (\ref{eq_binningta}).

\begin{figure}[ht]
\centering \includegraphics[angle=90,width=\columnwidth]{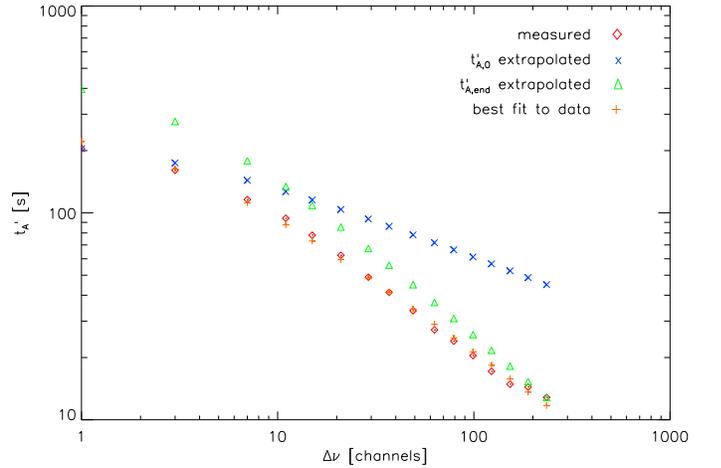}
\caption{Allan stability time $t\sub{A}'$ as a function of binning width
for the stability measurements shown in the lower plot of 
Fig. \ref{fig_binningdrift}. The red diamonds show the values measured
from the individual Allan variance spectra. The blue crosses and green triangles show
the theoretically expected behaviour for constant drift contributions
using $t\sub{A}'$ and $\alpha$ from the original (non-binned) data
or from the data with the largest binning width. The 
best fit to the measured data gives $t\sub{A}'=220$~s and $\alpha=1.4$.
}
\label{fig_tafit}
\end{figure}

Altogether, this {\changed makes} Eq. (\ref{eq_binningta}) 
questionable. Analysing the system tests of the HIFI instrument
in different configurations we noticed deviations in both
directions, cases with strong shallow noise, cases with narrow standing
waves easily suppressed by binning, and cases with a complete
cancellation, leading to an Allan time independent {\changed of} the binning
width. Finally we always performed the Allan variance analysis for
a number of binning widths. The result for the HIFI band 3a measurements
used in Fig. \ref{fig_binningdrift} is shown in Fig. \ref{fig_tafit}.
Together with the measured values of $t\sub{A}'(n\sub{bin})$ we
show the values that would be obtained from Eq. (\ref{eq_binningta})
using the Allan time and spectral drift coefficient $\alpha$ measured for the
first point (no binning) and the last point ($n\sub{bin}=235$) in the
plot. Both extrapolations overestimate the Allan stability time
at other binning widths.
We can, however, obtain a fit to the data using Eq. (\ref{eq_binningta})
by combining the information from all binning widths. This is shown
as violet crosses which provide a very good match to the measured
values. By using two different values for the exponent $\alpha$, one
for the drift slope at a particular binning width and one to
characterise the change of $t\sub{A}'$ with the binning size, we
can thus provide a full characterisation of the instrumental stability
with an accuracy of typically some ten percent.

\section{Optimisation of the observations}

\mbox{}\citet{SchiederKramer} used the Allan-variance minimum 
to derive constraints for the optimum timing of astronomical observations.
They considered drift contributions with the two spectral
indices $\alpha=2,3$ bracketing the behaviour often observed in 
spectroscopic Allan
variance measurements. In the analysis of numerous HIFI stability
measurements, we noticed, however, that a major fraction of total-power
Allan variance spectra and still a significant portion of spectroscopic
Allan variance spectra is dominated by shallower noise spectra,
often close to $1/f$ noise (see e.g. Fig. \ref{fig_tpaverage}). 
Thus we repeat their computation for
the general case of arbitrary noise spectral indices, also applying 
the new Allan time definition discussed above. Corresponding
equations for the Allan minimum time were also derived by Schieder
(priv. comm.).

Many observing modes are symmetric in the sense that equal time
intervals are spent on the astronomical source and on the reference.
These are beam-switch observations, moving either the telescope or
a fast chopping mirror between source and reference, and 
frequency-switch observations. To compensate linear drifts, the
most appropriate observing scheme consists of 
reference-source-source-reference sequences, with an integration time 
$t\sub{s}$ for each source phase, the same time $t\sub{s}$ in the reference 
phases, and a dead time $t\sub{d}$ in between. 

By using the information about the instrumental drift
obtained from the Allan variance spectrum, in particular the
Allan time $t\sub{A}'$ and the spectral index of the drift $\alpha$
it is possible to compute the average drift error in each difference
measurement $S_k-R_k$, where $S_k$ is the integrated
signal over the $k$th source phase and $R_k$ is the integrated signal
during the $k$th reference phase. The total uncertainty
of the astronomical measurement is then characterised by
the variance
\begin{equation}
\sigma\sub{obs}^2(t\sub{s},t\sub{d}) = \left \langle \left (
S_k-R_k - \left \langle S_k-R_k
\right \rangle_k \right)^2 \right \rangle_k \;.
\end{equation}

Following the formalism provided by \citet{SchiederKramer} we
can derive this variance normalised to the average signal level
from Eq. (\ref{eq_spectralallan})
\begin{eqnarray}
{\sigma\sub{obs}^2(t\sub{s},t\sub{d}) 
\over \langle s \rangle^2 }& =& {2 \over B\sub{Fl} t\sub{s}} \\
&&+ A { (2t\sub{s} +t\sub{d})^{\alpha+1} -2 (t\sub{s} +t\sub{d})^{\alpha+1}
+t\sub{d}^{\alpha+1}-2t\sub{s}^{\alpha+1} \over
2 (2^\alpha -2) t\sub{s}^2} \;. \nonumber
\label{eq_chopnoise}
\end{eqnarray}
The first term describes the radiometric noise which does not 
depend on the dead time. The second term is the drift noise. 
For $t\sub{d}=0$ we return to the known relation (\ref{eq_spectralallan})
for the Allan variance with no dead time between
subsequent data dumps. Eq. (\ref{eq_chopnoise}) holds for all spectral
indices $\alpha$ between 0 and 3 except for $\alpha=1$ where
a logarithmic divergence occurs leading to a somewhat different functional
description.

Substituting $A$ by $t\sub{A}'$ using Eq. (\ref{eq_atotarelation})
and normalising all times relative to the Allan time gives
\begin{eqnarray}
{\sigma\sub{obs}^2(x,d) \over \langle s \rangle^2 }
&=& {2 \over B\sub{Fl} t\sub{A}'} \\
&& \times \left( {1 \over x} +
{ (2x +d)^{\alpha+1} -2 (x +d)^{\alpha+1}
+d^{\alpha+1}-2x^{\alpha+1} \over
2 (2^\alpha -2) x^2} \right) \nonumber
\end{eqnarray}
with $x=t\sub{s}/t\sub{A}'$ and $d=t\sub{d}/t\sub{A}'$.  The use of the
traditional Allan minimum time in this equation would add a
factor $1/(\alpha-1)$ to the drift term. For $\alpha=2$ and
$\alpha=3$ one can then reproduce the {\changed results} obtained by
\citet{SchiederKramer}.

\begin{figure}[ht]
\centering \includegraphics[angle=90,width=\columnwidth]{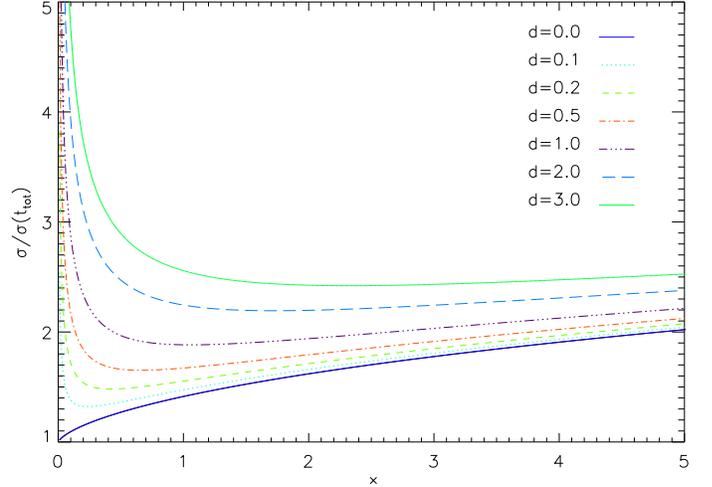}
\caption{Total noise in beam-switch observations consisting of
radiometric and drift contributions relative to the radiometric 
noise from an ideal instrument as a function of the relative chop phase
length $x$. The different curves 
represent different relative dead {\changed time} per cycle. A spectral index 
$\alpha=0.7$ is used for the drift noise.
\label{fig_sigmaplot}}
\end{figure}

The optimum observing mode is characterised by a minimum
total noise, composed of radiometric and drift noise,
obtained in a given observing time $t\sub{tot}$. Summing up
the $t\sub{tot} /(2t\sub{s}+t\sub{d})$ source-reference pairs fitting
in the total observing time, we obtain the total noise of the observation
as
\begin{eqnarray}
{\sigma\sub{tot}^2(x,d) \over \langle s \rangle^2 }
&=& {4x+2d \over B\sub{Fl} t\sub{tot}} \\
&& \times \left( {1 \over x} +
{ (2x +d)^{\alpha+1} -2 (x +d)^{\alpha+1}
+d^{\alpha+1}-2x^{\alpha+1} \over
2 (2^\alpha -2) x^2} \right) \nonumber
\label{eq_totalnoise}
\end{eqnarray}
(see Eq. \ref{eq_poisson}). The behaviour of this 
total noise is shown in Fig. \ref{fig_sigmaplot} for a drift spectral
index of 0.7 corresponding to the total-power measurements demonstrated
in Sect. \ref{sect_drifttoradio}. We plot the standard deviation
of the noise relative to the radiometric noise of an ideal instrument
having no dead times, i.e. relative to 
$\sigma\sub{ideal}/\langle s \rangle = 2/\sqrt{B\sub{Fl} t\sub{tot}}$,
as a function of the
integration time per cycle relative to the Allan time $t\sub{A}'$
for different relative dead times. We find the same general shape
of the curve as shown by \citet{SchiederKramer} for a drift spectral
index of 2, with a high noise for short integration times due to the
overhead from the dead time, an elevated noise for large integration times
due to the instrumental drift and a minimum defining the optimum 
integration time. However, for the shallow drift index, the minima 
are much wider than the corresponding minima computed by 
\citet{SchiederKramer}.

\begin{figure}[ht]
\centering \includegraphics[angle=90,width=\columnwidth]{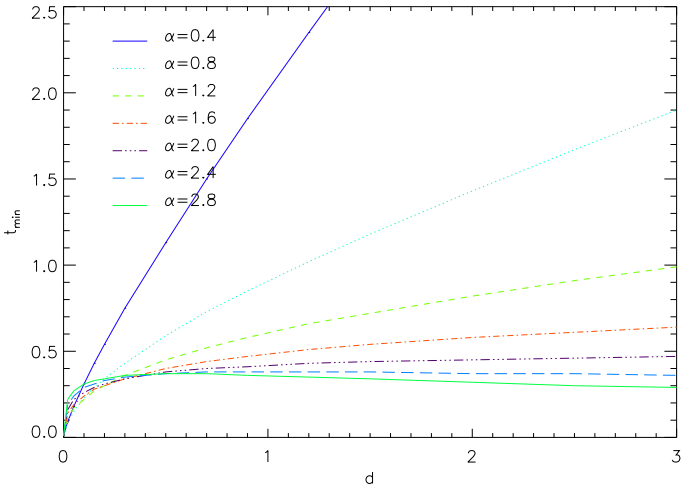}
\caption{The optimum phase length {\changed as a function of} the 
relative overhead from the dead time, {\changed $d=t\sub{d}/t\sub{A}'$,
for different spectral indices $\alpha$ of the instrumental drift.}}
\label{fig_tminplot}
\end{figure}

We can use this plot to compute optimum
cycle times from $t\sub{A}'$ in this case even if the Allan variance
spectrum shows no minimum. We obtain the optimum integration time 
per cycle as a function of the instrumental dead time and
the spectral index of the drift contributions by
\begin{eqnarray}
x\sub{opt}=t_{{\rm A},\Delta\nu}\!\!& {\rm root}_x\!\! & \!\! \left\{
(2x+d)^{\alpha+1}(\alpha x-d) + (x+d)^{\alpha+1}d 
\right.\nonumber\\
&& \left.  - 2(x+d)^{\alpha+1}(2x+d) (\alpha x+d) -d^{\alpha+1}(x+d)
 \right.\nonumber\\
&& \left.  -x^{\alpha+1}[\alpha(2x+d)-d] -(2^\alpha -2)xd
\right\}
\end{eqnarray}
where root$_x\{\}$ denotes {\changed the solution of the expression 
with respect to $x$.}

Fig. \ref{fig_tminplot} shows this optimum integration time
providing the minimum total noise as a function of the dead
time per cycle for a set of different spectral drift indices $\alpha$. 
For spectral indices between 1.5 and 3 the curves hardly
depend on the exact value of $\alpha$. When the dead time exceeds
half the Allan time the optimum integration time saturates also
at about half the Allan time. For shallow fluctuation spectra,
however, the optimum integration time increases {\changed rapidly} with
the dead time so that it can easily exceed the Allan time for
long dead times and spectral indices $\alpha<1$.

This effect somewhat relaxes the constraints for planning 
observations aimed at an accurate measurement of the
continuum level. Although the total-power Allan time $t\sub{A}'$
is usually very short, the corresponding spectral index
of the fluctuations it is often very shallow, $\alpha\approx 1$,
allowing cycle times which exceed the Allan time so that the
frequency for switching between source and reference can be
lower than the pure Allan time suggests.

\begin{figure}[ht]
\centering \includegraphics[angle=90,width=\columnwidth]{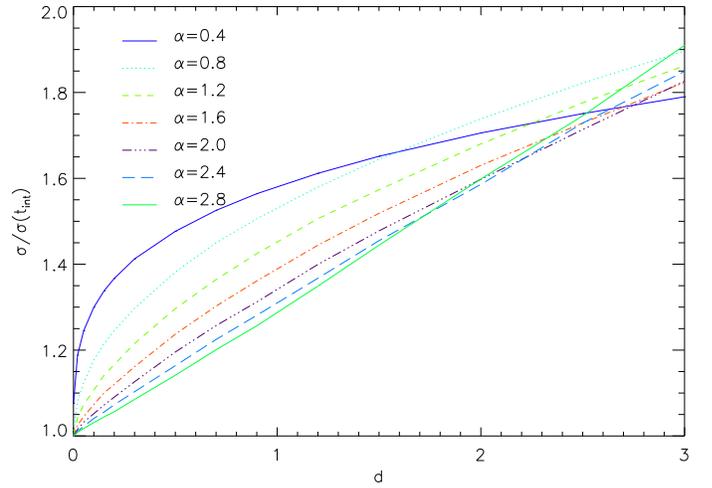}
\caption{Total noise relative to the radiometric noise for the optimum
phase length as a function of the relative dead time per cycle, {\changed  
$d=t\sub{d}/t\sub{A}'$,
for different spectral indices $\alpha$ of the instrumental drift.}}
\label{fig_relativerms}
\end{figure}

If the observations are set up with an optimum cycle length
we can use Eq. (\ref{eq_totalnoise}) to compare the expected drift 
noise from an observation with the radiometric noise. This ratio is
essential for the observer because it characterises the amplitude
of systematic baseline distortions due to instrumental fluctuations
relative to the white noise in the spectra. It is a measure {\changed of the} 
size of possible baseline ripples which might be visible in the
noise, {\changed and thus directly determine} the quality of the observations.
Figure \ref{fig_relativerms} shows the total noise relative
to the radiometric contribution for the optimum cycle length, i.e.
$\sigma\sub{tot,opt}/ \langle s \rangle \times \sqrt{B\sub{Fl} t\sub{tot}
/(4+2d/x)}$. In case of shallow drift spectra we notice a major drift
contribution already for relatively small dead times whereas steep 
fluctuation spectra result in an almost linear growth of the drift 
contribution with dead time. However, as long as the dead time  is 
smaller than about three Allan times, the total noise is increased
by less than a factor 2, i.e. the drift contribution can still be 
hidden in the radiometric noise of the baseline.

\begin{figure}[ht]
\centering \includegraphics[angle=90,width=\columnwidth]{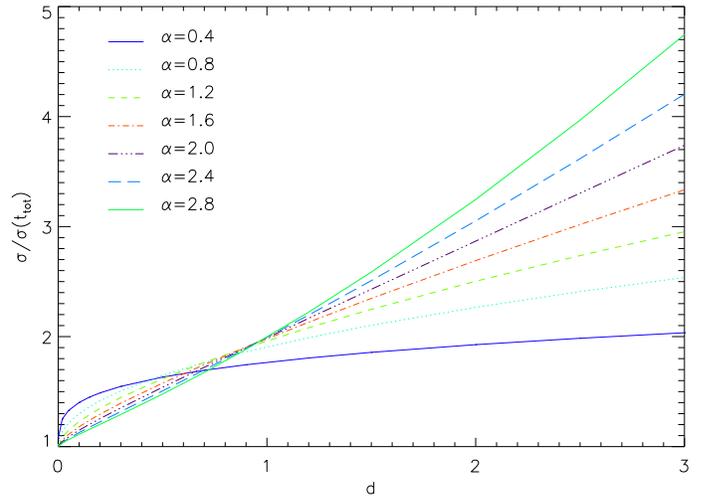}
\caption{Total noise RMS in {\changed a switched observation} relative
to the radiometric noise from an ideal instrument,
$\sigma\sub{tot,opt} \times \sqrt{B\sub{Fl} t\sub{tot}}
/ (2\langle s \rangle)$, for the optimum
phase length as a function of the relative dead time per cycle.}
\label{fig_totalrms}
\end{figure}

Finally we can use  Eq. (\ref{eq_totalnoise}) in astronomical
time estimates, to provide the user with realistic numbers for
the total data uncertainty obtainable in a given observing time.
Figure \ref{fig_totalrms} shows the total noise relative to
the radiometric noise from an ideal instrument, 
$2/\sqrt{B\sub{Fl} t\sub{tot}}$, for the optimum cycle
as a function of the spectral index of the instrumental drift 
and relative dead time.
For dead times exceeding about half the Allan time the 
drift contribution has the smallest impact for very shallow
noise spectra whereas steep spectra naturally create drift
contributions that grow quickly in time. The figure gives
a direct measure for the efficiency gain that can be obtained by
constructing a faster chopper {\changed mechanism or a more rapidly}
moving telescope
to minimise the dead time relative to the Allan time.

We can apply exactly the same approach to determine the timing
in second order observing loops set up to correct residuals from
the simple differencing measurements. A typical example {\changed is a} dual
beam-switch measurement where a chopping mirror switches quickly
between source and reference, but where the difference spectrum
contains some standing wave residuals due to the different
optical paths between the two mirror positions. {\changed
In this situation, a second loop is used} to periodically move the 
whole telescope so that the assignment of the two mirror positions to source
and reference is reversed. Adding {\changed the} differences from the two
telescope positions then removes the baseline residual while
maintaining the signal. To determine the period for the telescope
motions, we have to consider the stability of the baseline ripple,
i.e. the stability of the difference spectra. We call this
second-order or differential stability, because first order
instabilities will be removed already by the {\changed fast} reference
cycles. All computations used so far can be applied in the same
way to determine the period of the second-order loop if the
Allan variance analysis is not applied to continuous measurements,
but to differential measurements switching between source and 
reference where the normalised difference spectrum is used for 
$s_i(t_k)$. However, it is clear that the switching will 
introduce some dead times, so that the radiometric contribution
has to be corrected following Eq. (\ref{eq_deadtimes}). In our
HIFI tests, we found a very good second order stability for
dual-beam switch measurements under most conditions, so that
we could often only derive lower limits for the differential Allan
time. The second-order stability is considerably worse for
frequency-switch measurements following the same scheme.

The formalism described above has been implemented in HSPOT, the
Herschel observation planning tool, for the HIFI instrument
guaranteeing an optimum setup of the observations and providing
a realistic estimate for the quality of the data that can be
obtained in a given observing time. We have to stress, however, 
that it is applicable in the same way for ground based-telescopes
provided that stability measurements are performed for typical
atmospheric conditions to obtain estimates for the Allan time
and the drift exponent.

\section{Conclusions}

\subsection{The optimum method}

We propose a new scheme for the computation of the 
Allan variance of a time series of spectrometer data. It combines
the advantages of the spectroscopic Allan variance by
\citet{Schieder} with the advantages of the baseline
Allan variance by \citet{Siebertz} so that the same formalism
can be used to analyse the stability of an instrument with respect
to total-power fluctuations and with respect to spectroscopic
fluctuations.

We give two possible implementations for the algorithm to
compute the Allan variance spectra which differ in
the required computing power and the subjective ``smoothness''
of the resulting spectra  although {\changed both are} accurate within
the achievable uncertainty of the total Allan variance analysis.  
Although it is usually required to characterise an 
instrument by a single stability number, we show that 
the use of an average is not always justified, but has to be 
checked in each case by visualising the channel-by channel Allan
variance. For astronomical applications we propose the
``grand average'' subtraction scheme providing numbers
most relevant for astronomical line observations.

By introducing a new definition of the instrument stability 
time we can 
to characterise the instrument also in case of $1/f$ noise or
shallower fluctuation spectra at the cost of not being
directly comparable to the traditional Allan minimum time.
Using this stability time we compute the optimum observing
strategy in symmetric differencing observations for arbitrary 
drift parameters. If the appropriate loop timing is used, 
the total uncertainty of the measured spectra per given 
observing time will be minimised. In case of moderate
instrumental dead times relative to the Allan stability time 
it can guarantee that baseline ripples due to spectroscopic
drifts remain hidden in the radiometric noise.
If the instrument does not allow {\changed the use of} the optimum chop cycle
the drift noise can exceed the radiometric noise by
orders of magnitude.
The formalism {\changed presented here allows optimization 
of} both total-power and 
spectroscopic observations, but the considerably lower
total-power stability of any instrument {\changed demands} separate
setups for the two different scientific goals.

\subsection{Experience gained from testing HIFI}

The analysis of stability tests of the HIFI instrument showed
that gain fluctuations represent the main cause of instrumental
instabilities. We found that the temperature and mechanical 
stability is absolutely critical for a good performance.
It turned out that
changes in the pump level due to fluctuations in the LO power
or standing waves in the optical path between mixer and LO
are a {\changed major} cause of bad Allan stability times as predicted
by \citet{LO-instability}. 

This results in first instance in total power instabilities.
The total-power Allan times under good conditions can {\changed range from}
a few seconds up to about 20~s at 2~MHz fluctuation bandwidth.
Due to slight impedance mismatches between
mixer and amplifiers, {\changed any} change of the pump level also creates
a small change in the spectral response affecting the spectroscopic
stability. The standing waves in the LO path also show a characteristic
spectral instability imprint. Thus we obtain a reduced spectroscopic
stability as second-order byproduct of the gain fluctuations. 
Under good conditions we measured spectroscopic Allan times 
exceeding 100~s at 2~MHz fluctuation bandwidth. When removing
first order drifts by difference measurements, the differential
Allan variance of the resulting difference spectra measures
higher-order instabilities. {\changed The available differential schemes}
result in spectroscopic Allan
times of up to a few hundred seconds governing the period of second-order 
observational loops.

It turns out that spectroscopic binning removes part of the
fluctuations seen in the Allan variance spectra, but it enhances
the relative contribution of the underlying $1/f$ noise always present
in the gain fluctuations. Consequently, the impact of binning on the
Allan time cannot be predicted just from the measured Allan variance
spectrum at the native resolution of the instrument. It needs
to be measured, but can be fitted by two independent parameters.

The HIFI {\changed experience} can be used in the same way for ground based
telescope when the impact of the atmosphere is included
in the Allan variance measurements, i.e. if the stability
measurements are performed towards a celestial OFF position.
Fluctuations of the atmospheric transmission will again
affect mainly the total-power stability, but leave a characteristic
spectroscopic pattern in case of narrow absorption features
in the {\changed observed} spectral range. In case of a well-known shape
of the transmission function, an appropriate atmospheric model 
\citep[e.g.][]{ATM} may be used to correct for this spectroscopic 
instability.

\begin{acknowledgements}
I want to thank Rudolf Schieder, Jacob Kooi, and Oliver Siebertz 
for direct help in the computations and many useful discussions. 
Special thanks go to Paul Goldsmith for carefully reviewing the
manuscript providing many useful suggestions. This
work was supported by the DLR grant 50 OF 0006.
\end{acknowledgements}

\end{document}